\documentclass[preprint,10pt]{elsarticle}
\pdfoutput=1
\pdfcatalog{/PageLayout /SinglePage}



\makeatletter
\def\ps@pprintTitle{%
   \let\@oddhead\@empty
   \let\@evenhead\@empty
   \def\@oddfoot{\reset@font\hfil\thepage\hfil}
   \let\@evenfoot\@oddfoot
}
\makeatother

\usepackage{titlesec}
\usepackage{xcolor}
\usepackage{enumitem}
\usepackage{hyperref}
\hypersetup{colorlinks=true,urlcolor=black,pdfpagemode=UseNone,pdfstartview=Fit}
\usepackage[T1]{fontenc}\usepackage{lmodern}

\usepackage{algorithm}
\usepackage{algorithmic}

\usepackage{mathtools,amsfonts,amssymb}
\usepackage{caption}
\usepackage{subcaption}
\usepackage{booktabs}
\usepackage{pifont}
\usepackage{placeins}

\usepackage{float}

\newcommand{\pz}{\phantom{0}}

\newcommand{\pc}{\phantom{,}}
\newcommand{\vpad}{\vphantom{\bigg(}}

\DeclareMathOperator{\var}{Var}

\DeclareMathOperator{\e}{E}

\definecolor{orange}{rgb}{1,0.5,0}
\definecolor{green}{rgb}{0,0.5,0}
\definecolor{purple}{rgb}{0.5,0,0.5}
\newcommand{\reviewerOne}[1]{#1}
\newcommand{\rereading}[1]{#1}
\newcommand{\reviewerTwo}[1]{#1}

\usepackage[margin=1in]{geometry}
\biboptions{sort&compress}

\begin{document}

\begin{frontmatter}
\title{Code-Verification Techniques for Particle-in-Cell Simulations with Direct Simulation Monte Carlo Collisions}

\author[freno]{Brian A.\ Freno}
\ead{bafreno@sandia.gov}
\author[freno]{William J.\ McDoniel}
\author[freno]{Christopher H.\ Moore}
\author[freno]{Neil R.\ Matula}
\address[freno]{Sandia National Laboratories, Albuquerque, NM 87185}

\begin{abstract}
Particle-in-cell methods with stochastic collision models are commonly used to simulate collisional plasma dynamics, with applications ranging from hypersonic flight to semiconductor manufacturing.  Code verification of such methods is challenging due to the interaction between the spatial- and temporal-discretization errors, the statistical sampling noise, and the stochastic nature of the collision algorithm.  In this paper, we introduce our code-verification approaches to apply the method of manufactured solutions to plasma dynamics, and we derive expected convergence rates for the different sources of discretization and statistical error.  For the particles, we incorporate the method of manufactured solutions into the equations of motion.  We manufacture the particle distribution function and inversely query the cumulative distribution function to obtain known particle positions and velocities at each time step.  In doing so, we avoid modifying the particle weights, eliminating risks from potentially negative weights or modifications to weight-dependent collision algorithms.  For the collision algorithm, we average independent outcomes at each time step and we derive a corresponding manufactured source term for the velocity change for each particle.  By having known solutions for the particle positions and velocities, we are able to compute the error in these quantities directly instead of attempting to compute differences in distribution functions.  These approaches are equally valid for particle-in-cell simulations with Monte Carlo collisions and direct simulation Monte Carlo simulations of neutral gas flows.  We demonstrate the effectiveness of our approaches in three dimensions for different couplings between the particles and field, with and without binary elastic collisions, and with and without coding errors.
\end{abstract}


\end{frontmatter}

\section{Introduction} 
\label{sec:introduction}

Collisional plasma dynamics is relevant to many scientific and engineering applications.  For example, collisions among electrons, ions, and neutral species induce convective and radiative heat loads on vehicles in hypersonic flight and atmospheric reentry.  In plasma-based devices, such as lightning arresters and plasma switches, the breakdown voltage, conduction efficiency, and turn-off/on speed are extremely sensitive to the collision dynamics.
In pulsed-power systems, collisions govern high-energy-density physics as well as any potential gap closure that shunts the power away from the target.
In semiconductor manufacturing, collisions affect the energy distribution functions and thus etching and thin-film deposition.

The particle-in-cell method~\cite{harlow_1964,dawson_1983,birdsall_2004} is one of the most widely used techniques for simulating such phenomena.  Through this method, computational particles are placed randomly in phase space according to an initial distribution function.  The particle charge is interpolated onto the spatial mesh.  Maxwell's equations are solved on the spatial mesh for the electromagnetic field, which is interpolated onto the particles.  For each particle, the equations of motion due to the Lorentz-force acceleration~\cite{birdsall_2004} and particle collisions, modeled by Monte Carlo collisions~\cite{vahedi_1995} or direct simulation Monte Carlo~\cite{bird_1994}, are integrated.

\reviewerOne{%
In this work, we focus on binary elastic collisions modeled by direct simulation Monte Carlo.  This model is directly applicable to rarefied neutral-gas flows, weakly ionized gases, and collision models with finite total cross sections, including many particle--neutral interactions.  For charged-particle Coulomb collisions, additional extensions to collision models, such as the Takizuka--Abe or Nanbu models, are needed; however, such extensions are outside the scope of this work.}

Code verification is a necessary step towards establishing the credibility of computational physics simulations, with a focus on assessing the correctness of the implementation of the underlying numerical methods~\cite{roache_1998,knupp_2022,oberkampf_2010}.  For discretized equations, the most rigorous code-verification activities measure how the discretization error decreases with refinement.  Since exact solutions are known only for limited cases, the method of manufactured solutions~\cite{roache_2001} provides a means through which solutions can be constructed for problems of arbitrary complexity, fully or selectively exercising various code capabilities.

Code-verification activities have been demonstrated for many computational physics disciplines, including 
fluid dynamics and heat transfer~\cite{roy_2004,rader_2006,bond_2007,mcclarren_2008,veluri_2010,amar_2011,oliver_2012,edwards_2012,veeraragavan_2016,eca_2016,hennink_2021,freno_2021,freno_ablation_2022,bukreev_2023},
solid mechanics and fluid--structure interaction~\cite{chamberland_2010,etienne_2012,bukac_2023,fumagalli_2024},
electromagnetics~\cite{marchand_2013,marchand_2014,freno_em_mms_2020,amormartin_2021,freno_em_mms_quad_2021,freno_mfie_2022,freno_cfie_2023},
and plasma dynamics~\cite{turner_2016,riva_2017,banks_2019,tranquilli_2022,rueda_2023,medina_2023,rudi_2024,issan_2024}.  However, there are few examples for particle-in-cell methods.  

Particle-in-cell methods pose many code-verification challenges.  For example, in the electrostatic limit, in addition to the spatial-discretization error, the electric field incurs sampling and particle-position errors from the finite number of computational particles.  Likewise, in addition to the error from numerically integrating the equations of motion, the particles incur position and velocity errors due to the electric-field error sources and the stochastic collision algorithm, which operates on only a subset of particle pairs. 

Radtke et al.~\cite{radtke_2022} introduce a bootstrap-based power-law error model to differentiate between the stochastic noise and discretization error for code and solution verification.  Through this approach, the convergence rates are estimated with uncertainties, and code-verification examples are provided for simplified particle-in-cell problems with known solutions.
For multiple continuum-transport partial differential equations, Edwards and Bridson~\cite{edwards_2012} apply the method of manufactured solutions to assess how their high-order particle-in-cell scheme converges with respect to space and time for a fixed particle density.
For collisional plasma dynamics, Turner et al.~\cite{turner_2013, turner_2016} present a suite of benchmark verification tests consisting of unit and multi-unit tests that are based on analytic solutions.  From these tests, they demonstrate expected convergence rates.
Riva et al.~\cite{riva_2017} and Tranquilli et al.~\cite{tranquilli_2022} apply the method of manufactured solutions to collisionless plasma dynamics in one dimension for a single species~\cite{riva_2017} and in two dimensions for two species~\cite{tranquilli_2022} by manufacturing the particle distribution function and electric field.  For the particles, they modify the weight evolution equation using the Vlasov equation.  Consequently, the weighted empirical measure converges to the manufactured distribution function but the particles move independently of the manufactured distribution function.  Riva et al.~\cite{riva_2017} run several simulations for each discretization and provide multiple approaches with varying computational expense for measuring the error in the distribution function.  On the other hand, Tranquilli et al.~\cite{tranquilli_2022} argue that it is sufficient to measure the convergence of the grid quantities with a single simulation for each discretization.

In this paper, we present our code-verification approaches for three-dimensional particle-in-cell simulations with and without collisions, and we derive expected convergence rates.  For the particles, we incorporate the method of manufactured solutions into the equations of motion.  We manufacture the particle distribution function and inversely query the cumulative distribution function to obtain particle positions and velocities at each time step.  In doing so, we avoid modifying the weights, avoiding any risks from potentially negative weights or modifications to weight-dependent collision algorithms.  
For the collision algorithm, we average independent outcomes at each time step and we derive a corresponding manufactured source term for the velocity change for each particle.  To obtain a closed-form expression for this source term, we manufacture the cross section and anisotropy.
By having known solutions for the particle positions and velocities, we are able to compute the error in these quantities directly instead of attempting to compute differences in distribution functions, and we run one simulation per discretization.  In addition to measuring the convergence of the particle positions and velocities, we measure convergence in the scattering-angle distributions as a supplementary method for detecting coding errors.

This paper is organized as follows.  In Section~\ref{sec:equations}, we briefly discuss the particle equations of motion, the Poisson equation for the electric potential, and the particle-in-cell method.  In Section~\ref{sec:mms}, we describe our approach to applying the method of manufactured solutions to particle-in-cell simulations.  In Section~\ref{sec:error}, we derive expected convergence rates for the different sources of discretization and statistical error for use in our convergence studies.  In Section~\ref{sec:results}, we demonstrate the effectiveness of our approaches for different couplings between the particles and field, with and without collisions, and with and without coding errors.  In Section~\ref{sec:conclusions}, we summarize our work.

\section{Governing Equations} 
\label{sec:equations}

In this work, we consider a single-species collisional gas \reviewerTwo{on a spatially periodic domain} that interacts with an electric field.  The gas is represented by $N_p$ computational particles, and the collisions are binary and elastic.

The equations of motion for each computational particle $p\in[1,\,N_p]$ are
\begin{align}
\dot{w}_p(t) = 0, \qquad
\dot{\mathbf{x}}_p(t) = \mathbf{v}_p(t), \qquad 
\dot{\mathbf{v}}_p(t) = \frac{\mathbf{F}_p(t)}{m} + \biggl(\frac{\Delta \mathbf{v}_p(t)}{\Delta t}\biggr)_\text{coll},
\label{eq:eom_full}
\end{align}
where
$w_p$ denotes the weight of the computational particle (i.e., the number of physical particles it represents); 
$\mathbf{x}_p$ and $\mathbf{v}_p$ denote the position and velocity of the computational particle; 
$\mathbf{F}_p(t)=q\mathbf{E}_p(t)$ is the electrostatic Lorentz force, where $\mathbf{E}$ is the electric field and $\mathbf{E}_p(t)=\mathbf{E}(\mathbf{x}_p(t),t)$; 
$m$ and $q$ are the mass and charge of the particle species;
and $(\Delta \mathbf{v}_p/\Delta t)_\text{coll}$ denotes the instantaneous change in velocity over time $\Delta t$ due to a stochastic collision algorithm, such as the one described in Algorithm~\ref{alg:coll}.

When the influence of the magnetic field is negligible, the electric field can be expressed in terms of an electric potential $\phi$, where 
\begin{align}
\mathbf{E}=-\nabla\phi,
\label{eq:potential}
\end{align}
and Maxwell's equations reduce to the Poisson equation
\begin{align}
\Delta \phi = -\frac{\rho}{\epsilon_0},
\label{eq:poisson}
\end{align}
where 
\begin{align}
\rho(\mathbf{x},t)=q\int_{-\infty}^\infty f(\mathbf{x},\mathbf{v},t)d\mathbf{v}
\label{eq:charge_density}
\end{align}
is the charge density, $f(\mathbf{x},\mathbf{v},t)$ is the particle distribution function, and $\epsilon_0$ is the permittivity of free space.  \reviewerTwo{When solving~\eqref{eq:poisson} on the periodic domain, we subtract the spatial mean from the source term to obtain a well-posed problem, and we impose a zero-mean solution to ensure uniqueness.}

Through the particle-in-cell method, the weighted computational particles are placed randomly in phase space according to an initial distribution function.  The particle charge is interpolated onto the spatial mesh using the cloud-in-cell deposition scheme~\cite{birdsall_1969}.  The Poisson equation~\eqref{eq:poisson} is solved for the electric potential.  The electric field is computed from the potential and interpolated onto the particles.  For each particle, the equations of motion~\eqref{eq:eom_full} due to the Lorentz-force acceleration and the change in velocities due to the collisions are integrated.
\section{Manufactured Solutions} 
\label{sec:mms}

\reviewerOne{%
The method of manufactured solutions has been applied to collisionless particle-in-cell simulations in~\cite{riva_2017,tranquilli_2022}.  The authors manufacture the electric field and incorporate the corresponding manufactured source term into the Poisson equation.  For the particles, the authors manufacture the particle distribution function $f^M(\mathbf{x},\mathbf{v},t)$, then use the Vlasov equation
\begin{align}
\frac{\partial f}{\partial t} + \mathbf{v}\cdot\nabla_\mathbf{x} f + \frac{q}{m}\mathbf{E}
\cdot\nabla_\mathbf{v} f &{}= 0
\label{eq:vlasov}
\end{align}
to modify the weights of the computational particles so that the weighted empirical measure converges to the manufactured distribution function.  The particles move independently of the manufactured distribution function and their weights are adjusted accordingly.}  Consequently, this approach risks incurring negative weights or altering weight-dependent collision algorithms. To mitigate these risks, we instead apply the method of manufactured solutions to the equations of motion.

\subsection{Manufactured Particle Distribution Functions} 
We manufacture the particle distribution function as
\begin{align}
f^M(\mathbf{x},\mathbf{v},t) = f_\mathbf{x}(\mathbf{x},t) f_\mathbf{v}(\mathbf{v},t),
\label{eq:fm}
\end{align}
where
\begin{align}
f_{\mathbf{v}}(\mathbf{v},t) = \prod_{i=1}^3 f_{v_i}(v_i,t), 
\qquad 
f_{v_i}(v_i,t)=\frac{2}{\sqrt{\pi}}\frac{v_i^2}{\hat{v}_i(t)^3}e^{-v_i^2/\hat{v}_i(t)^2},
\qquad 
\int_{-\infty}^\infty f_{v_i}(v_i,t) dv_i = 1,
\label{eq:fv}
\end{align}
and
\begin{align}
f_\mathbf{x}(\mathbf{x},t) = N\prod_{i=1}^3 f_{x_i}(x_i,t),
\qquad
\int_{0}^{L_{x_i}} f_{x_i}(x_i,t) dx_i = 1,
\qquad
\int_V f_\mathbf{x}(\mathbf{x},t) d\mathbf{x} = N,
\label{eq:fx}
\end{align}
where $N$ is the number of physical particles in the volume $V=\prod_{i=1}^3 L_{x_i}$.  The forms of $f^M(\mathbf{x},\mathbf{v},t)$~\eqref{eq:fm} and the deliberately non-Maxwellian $f_{\mathbf{v}}(\mathbf{v},t)$~\eqref{eq:fv} are consistent with those used in~\cite{riva_2017,tranquilli_2022}; however, we have generalized $f_{\mathbf{v}}(\mathbf{v},t)$ to vary with time.  We impose the separability of $f_\mathbf{x}(\mathbf{x},t)$~\eqref{eq:fx} for convenience.

We additionally define cumulative distribution functions for each position component $x_i$ and each velocity component $v_i$:
\begin{align*}
F_{x_i}\bigl(x_i^M(t),t\bigr)=\int_{0}^{x_i^M(t)} f_{x_i}(x_i',t) dx_i',
\qquad
F_{v_i}\bigl(v_i^M(t),t\bigr)=\int_{-\infty}^{v_i^M(t)} f_{v_i}(v_i',t) dv_i'.
\end{align*}

\subsection{Manufactured Solutions through the Equations of Motion} 

We obtain our manufactured solutions $\mathbf{x}_p^M$ and $\mathbf{v}_p^M$ from the manufactured distribution function $f^M(\mathbf{x},\mathbf{v},t)$ \eqref{eq:fm}.  At the beginning of the simulation, we take $N_p$ uniform random samples $\xi_{x_{i_p}},\, \xi_{v_{i_p}}\in[0,\,1]$ for each $x_i$ and $v_i$, which we interpret as values of the initial cumulative distribution functions:
\begin{align*}
\xi_{x_{i_p}}=F_{x_i}\bigl(x_{i_p}^M(0),0\bigr), \qquad \xi_{v_{i_p}}=F_{v_i}\bigl(v_{i_p}^M(0),0\bigr).
\end{align*}
Throughout the simulation, we use these samples to inversely query the cumulative distribution functions to obtain the manufactured position components $x_{i_p}^M(t)$ and velocity components $v_{i_p}^M(t)$ for each computational particle at time $t$:
\begin{align}
F_{x_i}\bigl(x_{i_p}^M(t),t\bigr) = \xi_{x_{i_p}}, \qquad F_{v_i}\bigl(v_{i_p}^M(t),t\bigr) = \xi_{v_{i_p}}.
\label{eq:mms_sol}
\end{align}
\reviewerOne{By inversely querying the cumulative distribution functions, we obtain particle positions and velocities that are independent samples from the manufactured particle distribution.}

In general, since the distribution function dependencies are manufactured independently,  $\dot{\mathbf{x}}_p^M\ne\mathbf{v}_p^M$. We can differentiate~\eqref{eq:mms_sol} to compute $\dot{\mathbf{x}}_p^M$ and $\dot{\mathbf{v}}_p^M$:
\begin{align*}
\dot{x}_{i_p}^M(t) = - \frac{1}{\displaystyle f_{x_i}(x_i^M(t),t)}\displaystyle\int_{0}^{x_i^M(t)} \frac{\partial}{\partial t} f_{x_i}(x_i',t) dx_i', \qquad
\dot{v}_{i_p}^M(t) = - \frac{1}{\displaystyle f_{v_i}(v_i^M(t),t)}\displaystyle\int_{-\infty}^{v_i^M(t)} \frac{\partial}{\partial t} f_{v_i}(v_i',t) dv_i'.
\end{align*}
With $\mathbf{x}_p^M$, $\mathbf{v}_p^M$, $\dot{\mathbf{x}}_p^M$, and $\dot{\mathbf{v}}_p^M$ known, we modify the equations of motion~\eqref{eq:eom_full} as follows:
\begin{align}
\dot{\mathbf{x}}_p = \mathbf{v}_p + \dot{\mathbf{x}}_p^M - \mathbf{v}_p^M, 
\qquad 
\dot{\mathbf{v}}_p = \frac{q}{m}\mathbf{E}_p + \biggl(\frac{\Delta \mathbf{v}_p}{\Delta t}\biggr)_\text{coll} + \dot{\mathbf{v}}_p^M - \frac{q}{m}\mathbf{E}_p^M - \Biggl(\frac{\Delta \mathbf{v}_p^M}{\Delta t}\Biggr)_\text{coll},
\label{eq:eom_mms}
\end{align}
where $\mathbf{E}^M=-\nabla\phi^M$ is the manufactured electric field obtained from the manufactured electric potential $\phi^M$, and $\bigl(\Delta \mathbf{v}_p^M/\Delta t\bigr)_\text{coll}$ represents an analytic deterministic expression for the change due to collisions.  When we solve~\eqref{eq:eom_mms}, we compute the change in velocity due to the stochastic collision algorithm, $\bigl(\Delta \mathbf{v}_p\bigr)_\text{coll}$, as well as the change in velocity due to a corresponding manufactured source term, $\bigl(\Delta \mathbf{v}_p^M\bigr)_\text{coll}$. The evaluation of these terms is described in Section~\ref{sec:coll_src}.

\reviewerOne{%
In the absence of collisions, the Vlasov equation~\eqref{eq:vlasov} is effectively modified to be
\begin{align*}
\frac{\partial f}{\partial t} + \bigl(\mathbf{v}+\dot{\mathbf{x}}^M-\mathbf{v}^M\bigr)\cdot\nabla_\mathbf{x} f + \biggl(\frac{q}{m}\mathbf{E} + \dot{\mathbf{v}}^M-\frac{q}{m}\mathbf{E}^M\biggr)
\cdot\nabla_\mathbf{v} f &{}= 0,
\end{align*}
illustrating how the phase-space advection is modified.
}

\subsection{Manufactured Source Term for Binary Elastic Collisions} 
\label{sec:coll_src}

To demonstrate our approaches, we consider binary elastic collisions between same-species particles with uniform computational weights ($w_p=w=N/N_p$), modeled via the direct simulation Monte Carlo method.
Extensions to other collision models, such as Monte Carlo collisions or the Nanbu collision operator~\cite{nanbu_1997,higginson_2020}, and the inclusion of multispecies inelastic collisions between particles of varying weights are necessary to take full advantage of these approaches for many plasma codes; however, such extensions are outside the scope of this work.
%
%
Algorithm~\ref{alg:coll} provides an example collision algorithm for $\bigl(\Delta \mathbf{v}_p\bigr)_\text{coll}$.  
Such collision algorithms are stochastic and have an all-or-nothing effect on the particles; most particle velocities are unaffected, whereas a few change significantly in unpredictable directions.
On the other hand, for a manufactured-solution-type framework, the source terms, such as $\bigl(\Delta \mathbf{v}_p^M\bigr)_\text{coll}$, need to be deterministic.  
To mitigate this discrepancy, each time the collision algorithm would typically be queried, we instead query the collision algorithm $N_\text{avg}$ independent times and average the velocity change for each particle, replacing $\bigl(\Delta \mathbf{v}_p\bigr)_\text{coll}$ with 
\begin{align}
\bigl\langle\Delta \mathbf{v}_p\bigr\rangle_\text{coll} = \frac{1}{N_\text{avg}}\sum_{k=1}^{N_\text{avg}}\bigl(\Delta \mathbf{v}_p^k\bigr)_\text{coll},
\label{eq:avg_outcome}
\end{align}
and we derive the expected change in velocity for each particle: $\bigl(\Delta \mathbf{v}_p^M\bigr)_\text{coll}=\bigl\langle\Delta \mathbf{v}_p^M\bigr\rangle_\text{coll}$.

\subsubsection{Expected Change in Velocity for Each Particle} 

For binary elastic collisions, the post-collision velocities $\mathbf{v}'$ for colliding particles $p$ and $q$ are obtained from momentum and energy conservation:
\begin{align}
\mathbf{v}_p' = \frac{1}{2}(\mathbf{v}_q+\mathbf{v}_p + g\mathbf{n}),
\qquad
\mathbf{v}_q' = \frac{1}{2}(\mathbf{v}_q+\mathbf{v}_p - g\mathbf{n}),
\label{eq:vp_post}
\end{align}
where
\begin{align}
\mathbf{n} = \left\{\begin{matrix}\cos\epsilon\sin\chi \\ \sin\epsilon\sin\chi \\ \cos\chi\end{matrix}\right\}
\label{eq:n}
\end{align}
is the scattering direction unit vector, 
$\epsilon$ is the azimuthal scattering angle, $\chi$ is the polar scattering angle, and
$g = |\mathbf{v}_p-\mathbf{v}_q|$ is the magnitude of the relative velocity.
From~\eqref{eq:vp_post}, the change in velocity for particle $p$ is given by
\begin{align}
\Delta\mathbf{v}_p = \mathbf{v}_p' - \mathbf{v}_p = \frac{1}{2}(\mathbf{v}_q-\mathbf{v}_p + g\mathbf{n}).
\label{eq:delta_vp}
\end{align}

In a cell of volume $\Delta V$, the expected change in velocity for particle $p$ across possible collision partners is given by
\begin{align}
\bigl\langle\Delta \mathbf{v}_p^M\bigr\rangle_\text{coll}
&{}= \frac{\displaystyle\frac{N_p^\text{cell}-1}{2}\int_{\Delta V} \int_{-\infty}^\infty \int_0^{2\pi} \int_0^\pi P_\text{coll}(g)(\mathbf{v}_q-\mathbf{v}_p + g\mathbf{n})f^M(\mathbf{x}_q,\mathbf{v}_q,t)  p(\chi,\epsilon) d\chi d\epsilon d\mathbf{v}_q d\mathbf{x}_q}{\displaystyle\int_{\Delta V} \int_{-\infty}^\infty \int_0^{2\pi} \int_0^\pi f^M(\mathbf{x}_q,\mathbf{v}_q,t)  p(\chi,\epsilon) d\chi d\epsilon d\mathbf{v}_q d\mathbf{x}_q},
\label{eq:vel_change_1a}
\end{align}
where $p(\chi,\epsilon)$ is the joint probability density function for the scattering angles, with
\begin{align}
\int_0^{2\pi} \int_0^\pi p(\chi,\epsilon) d\chi d\epsilon=1,
\label{eq:p_unity}
\end{align}
and the per-pair collision probability over $\Delta t$ is
\begin{align}
P_\text{coll}(g) = \frac{\sigma(g) g w \Delta t }{\Delta V},
\label{eq:p_coll}
\end{align}
where $\sigma(g)$ is the total collision cross section.
We model the angular dependencies of the joint probability density function as separable: $p(\chi,\epsilon)=p_\chi(\chi)p_\epsilon(\epsilon)$, with
\begin{align}
\int_0^{2\pi} p_\epsilon(\epsilon) d\epsilon=1, \qquad
\int_0^\pi p_\chi(\chi) d\chi = 1. \label{eq:constraint_2}
\end{align}

When the scattering is azimuthally symmetric, 
\begin{align}
p_\epsilon(\epsilon)=\frac{1}{2\pi}, \qquad
F_{p_\epsilon}(\epsilon) = \int_0^\epsilon p_\epsilon(\epsilon') d\epsilon',
\label{eq:Fp_eps}
\end{align}
and $\epsilon=2\pi\xi_\epsilon$, where $\xi_\epsilon\in[0,\,1]$ is a uniform random sample.
The cumulative distribution function for $\chi$ is 
\begin{align}
F_{p_\chi}(\chi) = \int_0^\chi p_\chi(\chi') d\chi',
\label{eq:Fp_chi}
\end{align}
where, for another uniform random sample $\xi_\chi\in[0,\,1]$,
\begin{align}
\chi = F_{p_\chi}^{-1}(\xi_\chi).
\label{eq:chi}
\end{align}

In \eqref{eq:vel_change_1a}, 
\begin{align}
\int_0^{2\pi} \int_0^\pi \mathbf{n}p(\chi,\epsilon) d\chi d\epsilon=\frac{1}{2\pi}\int_0^{2\pi} \int_0^\pi \mathbf{n}p_\chi(\chi) d\chi d\epsilon= \biggl\{0,\, 0,\, \int_0^\pi p_\chi(\chi)\cos\chi d\chi\biggr\},
\label{eq:int_np}
\end{align}
such that, after accounting for~\eqref{eq:fm}, \eqref{eq:p_unity}, and~\eqref{eq:int_np},
\eqref{eq:vel_change_1a} becomes
\begin{align}
\bigl\langle\Delta \mathbf{v}_p^M\bigr\rangle_\text{coll}
&{}= \displaystyle\frac{w\Delta t\bigl(N_p^\text{cell}-1\bigr)}{2\Delta V} \int_{-\infty}^\infty 
\sigma(g) g f_\mathbf{v}(\mathbf{v}_q,t)\biggl[
(\mathbf{v}_q-\mathbf{v}_p )
+ g \biggl\{0,\, 0,\, \int_0^\pi p_\chi(\chi)\cos\chi d\chi\biggr\} \biggr]
 d\mathbf{v}_q.
\label{eq:vel_change_1b}
\end{align}

Equation~\eqref{eq:vel_change_1b} can be evaluated exactly if the integrand is a polynomial consisting of only even powers of $g$, including a constant term.  This can be accomplished by manufacturing $\sigma(g)$ to consist of odd powers of $g$ and by manufacturing $p_\chi(\chi)$ so that
\begin{align}
\int_0^\pi p_\chi(\chi)\cos\chi d\chi = 0
\label{eq:constraint_1}.
\end{align}
When~\eqref{eq:constraint_1} is satisfied, \eqref{eq:vel_change_1b} becomes
\begin{align}
\bigl\langle\Delta \mathbf{v}_p^M\bigr\rangle_\text{coll}
&{}= \frac{w\Delta t \bigl(N_p^\text{cell}-1\bigr)}{2 \Delta V}\int_{-\infty}^\infty \sigma(g) g(\mathbf{v}_q-\mathbf{v}_p)f_\mathbf{v}(\mathbf{v}_q,t) d\mathbf{v}_q.
\label{eq:vel_change_1c}
\end{align}

\subsubsection{Manufactured Anisotropy}

To satisfy~\eqref{eq:constraint_1}, we can consider isotropic scattering, where
\begin{align*}
p_\chi(\chi)=\frac{\sin\chi}{2}, \qquad F_{p_\chi}(\chi)=\frac{1-\cos\chi}{2},  \qquad \int_0^\pi p_\chi(\chi)\cos\chi d\chi=0,
\end{align*}
and $\chi$ is computed from~\eqref{eq:chi}:
\begin{align}
\chi = F_{p_\chi}^{-1}(\xi_\chi) = \cos^{-1}(1-2\xi_\chi).
\label{eq:chi_iso}
\end{align}

Alternatively, we can manufacture anisotropic scattering $p_\chi(\chi)\ne(\sin\chi)/2$ that permits an analytic expression for $\bigl\langle\Delta \mathbf{v}_p^M\bigr\rangle_\text{coll}$. 
To improve the likelihood of developing a closed-form expression for $F_{p_\chi}^{-1}$~\eqref{eq:chi}, we use the ansatz
\begin{align*}
p_\chi(\chi) = (C_0 + C_1\cos\chi + C_2\cos^2\chi+C_3\cos^3\chi)\sin\chi.
\end{align*}
Equation~\eqref{eq:constraint_1} is satisfied by $C_3=-5C_1/3$, and~\eqref{eq:constraint_2} is satisfied by $C_2=-3(2C_0-1)/2$, leading to
\begin{align}
p_\chi(\chi) = \biggl(C_0 + C_1\cos\chi  -\frac{3(2C_0-1)}{2}\cos^2\chi - \frac{5C_1}{3}\cos^3\chi\biggr)\sin\chi.
\label{eq:pchi_ms}
\end{align}
$C_0$ and $C_1$ are obtained by minimizing
\begin{align}
\int_0^\pi \bigl(p_\chi(\chi)-\bar{p}_\chi(\chi)\bigr)^2 d\chi,
\label{eq:anisotropy_opt}
\end{align}
where $\bar{p}_\chi(\chi)$ corresponds to a realistic anisotropic model.  

From~\eqref{eq:pchi_ms}, \eqref{eq:constraint_1} is satisfied and~\eqref{eq:vel_change_1c} is valid.

\subsubsection{Manufactured Cross Section}

If we manufacture the cross section using the form
\begin{align}
\sigma(g) = \sum_{n=0}^{\mathclap{N_\sigma-1}} \sigma_n g^{2n-1},
\label{eq:sigma_ms}
\end{align}
\eqref{eq:vel_change_1c} becomes
\begin{align}
\bigl\langle\Delta \mathbf{v}_p^M\bigr\rangle_\text{coll}
&{}= \frac{w\Delta t \bigl(N_p^\text{cell}-1\bigr)}{2 \Delta V}\sum_{n=0}^{N_\sigma-1} \sigma_n \mathbf{f}_n,
\label{eq:vel_change_3}
\end{align}
where
\begin{align}
\mathbf{f}_n = \int_{-\infty}^\infty g^{2n}(\mathbf{v}_q-\mathbf{v}_p)f_\mathbf{v}(\mathbf{v}_q,t) d\mathbf{v}_q
\label{eq:fn}
\end{align}
can be computed exactly.  For example, inserting~\eqref{eq:fv} into~\eqref{eq:fn} yields 
\begin{align*}
f_{i_0} &{}= -v_{i_p}, 
\\
f_{i_1} &{}= -\frac{1}{2}\bigl(3\hat{v}^2+6\hat{v}_i^2+2v_p^2\bigr)v_{i_p},
\\
f_{i_2} &{}= -\frac{1}{4}\biggl(15\hat{v}^4+36\hat{v}_i^4+24v_p^2\hat{v}_i^2+4v_p^4+24\hat{v}_i^2\hat{v}^2+12v_p^2\hat{v}^2+24\bigl(\hat{v}_1^2v_{1_p}^2+\hat{v}_2^2v_{2_p}^2+\hat{v}_3^2v_{3_p}^2\bigr)-12\prod_{j\ne i}\hat{v}_j^2\biggr)v_{i_p},
\end{align*}
where $\hat{v}^2=\sum_{i=1}^3 \hat{v}_i^2$.
%
%
%
%
%
%
%
%
%
%
%
%
%
%
%
%
%
%
Therefore, we have an analytic expression for $\bigl\langle\Delta \mathbf{v}_p^M\bigr\rangle_\text{coll}$~\eqref{eq:vel_change_3}.

A manufactured cross-section of the form in~\eqref{eq:sigma_ms} is more useful than a Maxwell-molecule cross section as it can be chosen to more closely resemble tabulated cross sections taken from experimental data, and it can more thoroughly test the code.

\subsection{Manufactured Solutions for the Poisson Equation} 

For the Poisson equation, we manufacture the electric potential $\phi^M(\mathbf{x},t)$ and modify~\eqref{eq:poisson} to be
\begin{align}
\Delta \phi = -\frac{\rho}{\epsilon_0} + \Delta \phi^M + \frac{\rho^M}{\epsilon_0},
\label{eq:poisson_mms}
\end{align}
where $\Delta \phi^M$ is evaluated analytically.  From~\eqref{eq:charge_density} and~\eqref{eq:fm},
\begin{align*}
\rho^M(\mathbf{x},t)=q\int_{-\infty}^\infty f^M(\mathbf{x},\mathbf{v},t)d\mathbf{v} = q f_\mathbf{x}(\mathbf{x},t)
\end{align*}
is evaluated analytically as well.

\reviewerOne{%
\subsection{Implementation Considerations} 

From an implementation perspective, our approach requires manufactured expressions for the particle distribution function dependencies, the electric potential, and the corresponding source terms, as well as the ability to inversely query the manufactured cumulative distribution functions to initialize and evaluate the manufactured particle positions and velocities.  When collisions are modeled, each collision-algorithm call is replaced by an average over $N_\text{avg}$ independent realizations together with the evaluation of the corresponding analytic source term.  These additions are localized primarily to the initialization, source-term evaluation, and diagnostic portions of the code, leaving the particle trajectory evolution, charge deposition, field solve, and basic collision algorithm largely unmodified.  Furthermore, our approach does not modify the particle weights.
}
\section{Error Analysis} 
\label{sec:error}

As described in the introduction, we measure the rate at which the numerical solutions converge to the manufactured solutions to assess the correctness of the implementation of the underlying methods.  In this section, we discuss the expected convergence rates.

To solve~\eqref{eq:eom_mms} and~\eqref{eq:poisson_mms}, we discretize the spatial domain with uniform cells of size $\Delta x_i$, we numerically integrate with time-step size $\Delta t$, we represent the physical particles with $N_p$ computational particles, and we run the collision algorithm $N_\text{avg}$ times and average the outcomes.  We refine these quantities together, such that
\begin{align}
\Delta x_i \sim\Delta t\sim h, 
\qquad
N_p\sim h^{-\reviewerTwo{s}},
\qquad
N_\text{avg}\sim h^{-r},
\qquad
N_\text{cell}\sim h^{-3}, 
\qquad
N_p^\text{cell}\sim h^{-(\reviewerTwo{s}-3)},
\label{eq:refinements}
\end{align}
where $\reviewerTwo{s}$ and $r$ are integer parameters in the refinement strategy that are chosen to be the minimum values necessary to achieve a given accuracy.  The smaller the values, the lower the computational expense.

\subsection{Field Quantities} 
\label{sec:error_field}

For the field quantities, the error in the Poisson equation~\eqref{eq:poisson} is due to the trilinear finite element basis functions, the $N_p$-particle sampling of the charge density, and the error in the particle positions.  
The basis-function error is $\mathcal{O}(h^2)$ for $\phi$, and we compute $\mathbf{E}$~\eqref{eq:potential} using a second-order-accurate gradient operator so that its discretization error is also $\mathcal{O}(h^2)$.
The sampling error in $\phi$ is $\mathcal{O}\bigl(N_p^{-1/2}\bigr)$, and the sampling error in $\mathbf{E}$ is $\mathcal{O}\bigl(N_p^{-1/2}h^{-1/2}\bigr)$~\cite{ricketson_2017,riva_2017,tranquilli_2022}. 
If the error in the particle positions is $\mathcal{O}\bigl(h^{p_\mathbf{x}}\bigr)$, the particle-position error for $\phi$ and $\mathbf{E}$ is, at worst, $\mathcal{O}(h^{p_\mathbf{x}})$.
Therefore, from~\eqref{eq:refinements}, the errors in $\phi$ and $\mathbf{E}$ are
\begin{alignat}{7}
e_\phi &{}= 
\mathcal{O}\bigl(h^2\bigr) + \mathcal{O}\bigl(N_p^{-1/2}\bigr) &&{}+ \mathcal{O}\bigl(h^{p_\mathbf{x}}\bigr) &&{}=
\mathcal{O}\bigl(h^2\bigr) + \mathcal{O}\bigl(h^{\reviewerTwo{s}/2}\bigr) &&{}+ \mathcal{O}\bigl(h^{p_\mathbf{x}}\bigr) &&{}= 
\mathcal{O}\bigl(h^{p_\phi}\bigr), \nonumber\\
\mathbf{e}_\mathbf{E} &{} = 
\mathcal{O}\bigl(h^2\bigr) + \mathcal{O}\bigl(N_p^{-1/2}h^{-1/2}\bigr) &&{}+ \mathcal{O}\bigl(h^{p_\mathbf{x}}\bigr) &&{}=
\mathcal{O}\bigl(h^2\bigr) + \mathcal{O}\bigl(h^{(\reviewerTwo{s}-1)/2}\bigr) &&{}+ \mathcal{O}\bigl(h^{p_\mathbf{x}}\bigr) &&{}=  
\mathcal{O}\bigl(h^{p_\mathbf{E}}\bigr), \label{eq:e_e}
\end{alignat}
where 
\begin{align}
p_\phi = \min\biggl\{2,\,\frac{\reviewerTwo{s}}{2},\,p_\mathbf{x}\biggr\}, \qquad
p_\mathbf{E} = \min\biggl\{2,\,\frac{\reviewerTwo{s}-1}{2},\,p_\mathbf{x}\biggr\},
\label{eq:p_phi}
\end{align}
and $p_\mathbf{x}=p_x=p_y=p_z$ is determined in Section~\ref{sec:error_accumulation}.

\subsection{Particles} 
\label{sec:error_particles}

For the particles, we integrate~\eqref{eq:eom_mms} in time using a second-order-accurate velocity-Verlet approach:
\begin{align}
\mathbf{v}_p^{n+1/2} &{}= \mathbf{v}_p^{n\phantom{{}+0/1}} + \frac{1}{2}\Biggl(\Delta t\frac{q}{m}\bigl(\mathbf{E}_p-\mathbf{E}_p^M\bigr)^{n\phantom{{}+0}} + \bigl\langle\Delta \mathbf{v}_p\bigr\rangle_\text{coll}^{n\phantom{{}+0/1}} - \bigl\langle\Delta \mathbf{v}_p^M\bigr\rangle_\text{coll}^{n\phantom{{}+0/1}} + \Delta t\bigl(\dot{\mathbf{v}}_p^M\bigr)^{n\phantom{{}+0}}\Biggr) + \boldsymbol{\tau}_{\mathbf{v}_p}^n, 
\nonumber\\[-.25em]
\mathbf{x}_p^{n+1\phantom{/1}} &{}= \mathbf{x}_p^{n\phantom{{}+0/1}} + \Delta t \Bigl(\mathbf{v}_p + \dot{\mathbf{x}}_p^M - \mathbf{v}_p^M\Bigr)^{n+1/2}      + \boldsymbol{\tau}_{\mathbf{x}_p}^n, 
\nonumber\\[-.25em]
\mathbf{v}_p^{n+1\phantom{/1}} &{}= \mathbf{v}_p^{n+1/2} + \frac{1}{2}\Biggl(\Delta t\frac{q}{m}\bigl(\mathbf{E}_p-\mathbf{E}_p^M\bigr)^{n+1} + \bigl\langle\Delta \mathbf{v}_p\bigr\rangle_\text{coll}^{n+1/2} - \bigl\langle\Delta \mathbf{v}_p^M\bigr\rangle_\text{coll}^{n+1/2} + \Delta t\bigl(\dot{\mathbf{v}}_p^M\bigr)^{n+1}\Biggr) + \boldsymbol{\tau}_{\mathbf{v}_p}^{n+1/2},
\label{eq:velocity_verlet}
\end{align}
where $\boldsymbol{\tau}$ denotes the truncation error of the time integration, with $\boldsymbol{\tau}_{\mathbf{v}_p}^n+\boldsymbol{\tau}_{\mathbf{v}_p}^{n+1/2}\sim \mathcal{O}(\Delta t^3)$ and $\boldsymbol{\tau}_{\mathbf{x}_p}^n\sim \mathcal{O}(\Delta t^3)$.
In addition to the per-step time-integration error, the particles incur the per-step Lorentz-force-acceleration error at time step $n$
\begin{align}
\mathbf{e}_{\text{acc}_p}^n = \Delta t \frac{q}{m}\bigl(\mathbf{E}_p-\mathbf{E}_p^M\bigr)^n,
\label{eq:e_acc}
\end{align}
and the per-step collision error
\begin{align}
\mathbf{e}_{\text{coll}_p}^n = \bigl\langle\Delta \mathbf{v}_p\bigr\rangle_\text{coll}^n - \bigl\langle\Delta \mathbf{v}_p^M\bigr\rangle_\text{coll}^n=
\frac{1}{N_\text{avg}} \sum_{k=1}^{N_\text{avg}}\bigl(\Delta \mathbf{v}_p^k\bigr)_\text{coll}^n 
- \bigl\langle\Delta \mathbf{v}_p^M\bigr\rangle_\text{coll}^n.
\label{eq:e_coll}
\end{align}

\subsubsection{Lorentz-Force Acceleration} 

The per-step Lorentz-force-acceleration error $\mathbf{e}_{\text{acc}_p}^n$~\eqref{eq:e_acc} is proportional to the product of $\mathbf{e}_\mathbf{E}$~\eqref{eq:e_e} and $\Delta t$, such that
\begin{align*}
\mathbf{e}_{\text{acc}_p}^n = \mathcal{O}\bigl(h^{p_\mathbf{E}}\Delta t\bigr) = \mathcal{O}\bigl(h^{p_\mathbf{E}+1}\bigr) = \mathcal{O}\bigl(h^{p_\text{acc}}\bigr),
\end{align*}
where
\begin{align*}
p_\text{acc} = \min\biggl\{3,\,\frac{\reviewerTwo{s}+1}{2},\,p_\mathbf{x}+1\biggr\}.
\end{align*}

\subsubsection{Collisions} 
\label{sec:collisions}

By construction, the expected collision error $\mathbf{e}_{\text{coll}_p}^n$~\eqref{eq:e_coll} is zero. The variance of $\mathbf{e}_{\text{coll}_p}^n$ arises from two independent sources of variance in $\langle\Delta \mathbf{v}_p\rangle_\text{coll}^n$: 
\begin{enumerate}
\item the $N_p^\text{cell}$-particle sampling of $f_\mathbf{v}(\mathbf{v},t)$ within the collision algorithm for each cell, and
\item the averaging of the $N_\text{avg}$ runs of the collision algorithm on those velocities.
\end{enumerate}
Let 
$\mathcal{V}={\{\mathbf{v}_q\}}_{q=1}^{N_p^\text{cell}}$ denote the $N_p^\text{cell}$ independent velocity samples from $f_\mathbf{v}(\mathbf{v},t)$, and
let
$\mathcal{K}=\{\Delta \mathbf{v}_p^k\}_{k=1}^{N_\text{avg}}$ denote the $N_\text{avg}$ changes in velocity from the independent collision-algorithm runs for a fixed $\mathcal{V}$.
We first average over collision-algorithm outcomes $\mathcal{K}$ for a fixed $\mathcal{V}$, then we average over all possible velocity sets $\mathcal{V}$.
Let
$\e_\mathcal{K}[{}\cdot{}|\mathcal{V}]$ and $\var_\mathcal{K}[{}\cdot{}|\mathcal{V}]$ denote the expectation and variance over the $N_\text{avg}$ collision-algorithm runs at fixed $\mathcal{V}$,
and 
$\e_\mathcal{V}[{}\cdot{}]$ and $\var_\mathcal{V}[{}\cdot{}]$ denote the expectation and variance over all choices of $\mathcal{V}$.
From the law of total variance, for velocity component $i$,
\begin{align}
\var\Bigl[\bigl\langle\Delta v_{i_p}\bigr\rangle_\text{coll}^n\Bigr] = 
  \e_\mathcal{V}\Bigl[\var_\mathcal{K}\Bigl[\bigl\langle\Delta v_{i_p}\bigr\rangle_\text{coll}^n \Bigm| \mathcal{V}\Bigr]\Bigr] + 
\var_\mathcal{V}\Bigl[  \e_\mathcal{K}\Bigl[\bigl\langle\Delta v_{i_p}\bigr\rangle_\text{coll}^n \Bigm| \mathcal{V}\Bigr]\Bigr].
\label{eq:total_variance}
\end{align}

In~\eqref{eq:total_variance}, the first term is the algorithmic variance, which is the variance of the change in velocity averaged over all possible velocity samples $\mathcal{V}$. From~\eqref{eq:avg_outcome},
\begin{align*}
\e_{\mathcal{V}}\Bigl[\var_\mathcal{K}\Bigl[\bigl\langle\Delta v_{i_p}\bigr\rangle_\text{coll}^n \Bigm| \mathcal{V}\Bigr]\Bigr] 
=\e_{\mathcal{V}}\Biggl[\var_\mathcal{K}\Biggl[    \frac{1}{N_\text{avg}} \sum_{k=1}^{N_\text{avg}}\bigl(\Delta v_{i_p}^k\bigr)_\text{coll}^n      \Bigm| \mathcal{V}\Biggr]\Biggr]
=\frac{1}{N_\text{avg}^2}\e_{\mathcal{V}}\Biggl[\var_\mathcal{K}\Biggl[     \sum_{k=1}^{N_\text{avg}}\bigl(\Delta v_{i_p}^k\bigr)_\text{coll}^n      \Bigm| \mathcal{V}\Biggr]\Biggr].
\end{align*}
Noting that the collision-algorithm runs are independent, 
\begin{align*}
\frac{1}{N_\text{avg}^2}\e_{\mathcal{V}}\Biggl[\var_\mathcal{K}\Biggl[     \sum_{k=1}^{N_\text{avg}}\bigl(\Delta v_{i_p}^k\bigr)_\text{coll}^n      \Bigm| \mathcal{V}\Biggr]\Biggr]
=
\frac{1}{N_\text{avg}^2}\e_{\mathcal{V}}\Biggl[    \sum_{k=1}^{N_\text{avg}}\var_\mathcal{K}\Bigl[ \bigl(\Delta v_{i_p}^k\bigr)_\text{coll}^n      \Bigm| \mathcal{V}\Bigr]\Biggr],
\end{align*}
and that each velocity change has the same conditional variance for a fixed $\mathcal{V}$,
\begin{align*}
\frac{1}{N_\text{avg}^2}\e_{\mathcal{V}}\Biggl[    \sum_{k=1}^{N_\text{avg}}\var_\mathcal{K}\Bigl[ \bigl(\Delta v_{i_p}^k\bigr)_\text{coll}^n      \Bigm| \mathcal{V}\Bigr]\Biggr]
=
\frac{1}{N_\text{avg}}\e_{\mathcal{V}}\bigl[   \var_\mathcal{K}\bigl[ \bigl(\Delta v_{i_p}\bigr)_\text{coll}^n      \bigm| \mathcal{V}\bigr]\bigr].
\end{align*}
From $P_\text{coll}$~\eqref{eq:p_coll}, the variance of the change in velocity is proportional to $\Delta t$, such that we can write
\begin{align*}
\e_{\mathcal{V}}\bigl[   \var_\mathcal{K}\bigl[ \bigl(\Delta v_{i_p}\bigr)_\text{coll}^n      \bigm| \mathcal{V}\bigr]\bigr] = C_{i_\text{alg}}^n\Delta t,
\end{align*}
where $C_{i_\text{alg}}^n$ is independent of $N_\text{avg}$, $N_p$, and $\Delta t$.  Consequently, in~\eqref{eq:total_variance},
\begin{align}
\e_{\mathcal{V}}\Bigl[\var_\mathcal{K}\Bigl[\bigl\langle\Delta v_{i_p}\bigr\rangle_\text{coll}^n \Bigm| \mathcal{V}\Bigr]\Bigr]  = \frac{C_{i_\text{alg}}^n\Delta t}{N_\text{avg}}.
\label{eq:alg_var}
\end{align}

The second term in~\eqref{eq:total_variance} is the sampling variance, which is the variance of the mean change in velocity over all possible velocity samples $\mathcal{V}$.  
Along the lines of~\eqref{eq:vel_change_1c}, we can write
\begin{align}
\bigl\langle\Delta v_{i_p}\bigr\rangle_\text{coll}
&{}= \frac{w\Delta t \bigl(N_p^\text{cell}-1\bigr)}{2 \Delta V} I_{i_p},
\label{eq:vel_change_1d}
\end{align}
where
\begin{align*}
I_{i_p} = \int_{-\infty}^\infty \sigma(g) g(v_{i_q}-v_{i_p})f_\mathbf{v}(\mathbf{v}_q,t) d\mathbf{v}_q.
\end{align*}
With $w=N/N_p$, $N_\text{cell}=V/\Delta V$, and $N_p^\text{cell}= C_{N_p^\text{cell}}^n N_p/N_\text{cell}$, then $w/\Delta V = C_{N_p^\text{cell}}^n N / (V N_p^\text{cell})$, and~\eqref{eq:vel_change_1d} can be written as 
\begin{align}
\bigl\langle\Delta v_{i_p}\bigr\rangle_\text{coll}
&{}= \frac{C_{N_p^\text{cell}}^n N\Delta t \bigl(N_p^\text{cell}-1\bigr)}{2 V N_p^\text{cell}} I_{i_p}.
\label{eq:vel_change_1e}
\end{align}
However, instead of evaluating $I_{i_p}$, we sample its integrand:
\begin{align}
\tilde{I}_{i_p} = \frac{1}{N_p^\text{cell}-1} \sum_{q\ne p} H_i(\mathbf{v}_p,\mathbf{v}_q),
\label{eq:I_tilde}
\end{align}
where
\begin{align*}
H_i(\mathbf{v}_p,\mathbf{v}_q) = \sigma(g)g(v_{i_q}-v_{i_p}),
\end{align*}
such that, from~\eqref{eq:vel_change_1e} and~\eqref{eq:I_tilde},
\begin{align}
\e_\mathcal{K}\Bigl[\bigl\langle\Delta v_{i_p}\bigr\rangle_\text{coll}^n \Bigm| \mathcal{V}\Bigr]
=
\frac{C_{N_p^\text{cell}}^n N\Delta t \bigl(N_p^\text{cell}-1\bigr)}{2 V N_p^\text{cell}} \tilde{I}_{i_p}
=
\frac{C_{N_p^\text{cell}}^n N\Delta t }{2 V N_p^\text{cell}}  \sum_{q\ne p} H_i(\mathbf{v}_p,\mathbf{v}_q).
\label{eq:samp_1}
\end{align}
Inserting~\eqref{eq:samp_1} into the second term in~\eqref{eq:total_variance},
\begin{align}
\var_\mathcal{V}\Bigl[  \e_\mathcal{K}\Bigl[\bigl\langle\Delta v_{i_p}\bigr\rangle_\text{coll}^n \Bigm| \mathcal{V}\Bigr]\Bigr] 
=
\Biggl(\frac{C_{N_p^\text{cell}}^n N\Delta t }{2 V N_p^\text{cell}}\Biggr)^2  \var_\mathcal{V}\Bigl[\sum_{q\ne p} H_i(\mathbf{v}_p,\mathbf{v}_q)\Bigr].
\label{eq:samp_var_1}
\end{align}
Since each $H_i(\mathbf{v}_p,\mathbf{v}_q)$ is independent and has the same conditional variance for a fixed $\mathcal{V}$,
\begin{align}
\var_\mathcal{V}\Bigl[\sum_{q\ne p} H_i(\mathbf{v}_p,\mathbf{v}_q)\Bigr] = 
\sum_{q\ne p} \var_\mathcal{V}\bigl[H_i(\mathbf{v}_p,\mathbf{v}_q)\bigr] =
(N_p^\text{cell}-1) \var_\mathcal{V}\bigl[H_i(\mathbf{v}_p,\mathbf{v})\bigr].
\label{eq:H_var}
\end{align}
Inserting~\eqref{eq:H_var} into~\eqref{eq:samp_var_1},
\begin{align}
\var_\mathcal{V}\Bigl[  \e_\mathcal{K}\Bigl[\bigl\langle\Delta v_{i_p}\bigr\rangle_\text{coll}^n \Bigm| \mathcal{V}\Bigr]\Bigr] 
=
\Biggl(\frac{C_{N_p^\text{cell}}^n N\Delta t }{2 V N_p^\text{cell}}\Biggr)^2  (N_p^\text{cell}-1) \var_\mathcal{V}\bigl[H_i(\mathbf{v}_p,\mathbf{v})\bigr]
\approx
\frac{C_{i_\text{samp}}^n (\Delta t)^2}{N_p^\text{cell}},
\label{eq:samp_var}
\end{align}
where $C_{i_\text{samp}}^n$ is independent of $N_\text{avg}$, $N_p$, and $\Delta t$.

Therefore, from~\eqref{eq:total_variance}, \eqref{eq:alg_var}, and~\eqref{eq:samp_var}, the variance of the per-step collision error is
\begin{align}
\var\Bigl[\bigl(\mathbf{e}_{\text{coll}_p}^n\bigr)_i\Bigr] = 
\var\Bigl[\bigl\langle\Delta v_{i_p}\bigr\rangle_\text{coll}^n\Bigr] = 
\frac{C_{i_\text{alg}}^n \Delta t   }{N_\text{avg}} + 
\frac{C_{i_\text{samp}}^n(\Delta t)^2}{N_p^\text{cell}}.
\label{eq:var}
\end{align}
We assume the per-step error is proportional to its standard deviation, such that, from~\eqref{eq:refinements} and~\eqref{eq:var}, the per-step collision error is
\begin{align}
\mathbf{e}_{\text{coll}_p}^n = \mathcal{O}\bigl(N_\text{avg}^{-1/2}(\Delta t)^{1/2}\bigr) + \mathcal{O}\Bigl({N_p^\text{cell}}^{-1/2}\Delta t\Bigr) 
=
\mathcal{O}\bigl(h^{(r+1)/2}\bigr) + \mathcal{O}\bigl(h^{(\reviewerTwo{s}-1)/2}\bigr) = \mathcal{O}\bigl(h^{p_\text{coll}}\bigr),
\label{eq:e_coll_pn}
\end{align}
where
\begin{align*}
p_\text{coll} = \min\biggl\{\frac{r+1}{2},\,\frac{\reviewerTwo{s}-1}{2}\biggr\},
\end{align*}
and that $\mathbf{e}_{\text{coll}_p}^n$ remains sufficiently small so that all inter-step covariances are higher order and therefore negligible.  Equation~\eqref{eq:e_coll_pn} shows that, if $N_\text{avg}$ is fixed ($r=0$), the leading term of $\mathbf{e}_{\text{coll}_p}^n$ is, at best, $\mathcal{O}(h^{1/2})$.  As described in Section~\ref{sec:error_accumulation}, the accumulation of such a local error does not converge.

This analysis additionally assumes that the collision events are independently obtained from all possible pairs.  For a given cell, there are $N_\text{poss}=N_p^\text{cell}(N_p^\text{cell}-1)/2$ possible pairs.  At a given time step, there are $N_\text{events}=N_\text{avg} N_\text{pairs}$ collision events, where $N_\text{pairs}= N_\text{poss} P_{\text{coll}}(g) \le N_\text{poss} P_{\text{coll}_\text{max}}$.  For the assumption to hold, we require $N_\text{events}\ll N_\text{poss}$, such that, from~\eqref{eq:p_coll} a sufficient condition is
\begin{align}
N_\text{avg} \ll \frac{1}{P_{\text{coll}_\text{max}}} = \frac{V}{ N(\sigma g)_\text{max}} \frac{N_p^\text{cell}}{ \Delta t}.
\label{eq:navg_ineq}
\end{align}

\subsection{Error Accumulation} 
\label{sec:error_accumulation}

We are interested in the following discrete error norms at $t=T=N_{\Delta t}\Delta t$ over all $N_p$ particles in the domain: 
\begin{align*}
\varepsilon_2^{\alpha} = \sqrt{\frac{1}{N_p} \sum_{p=1}^{N_p}\Bigl(e_{\alpha_p}^{N_{\Delta t}}\Bigr)^2}=\mathcal{O}\bigl(h^{p_\alpha}\bigr), 
\qquad
\varepsilon_\infty^{\alpha} = \max_p \bigl|e_{\alpha_p}^{N_{\Delta t}}\bigr|=\mathcal{O}\bigl(h^{p_\alpha}\bigr), 
\end{align*}
where $e_{\alpha_p} = \alpha_p - \alpha_p^M$,
and $\alpha=\{x,y,z,u,v,w\}$.  

As stated in Section~\ref{sec:error_particles}, the local errors in the particle positions and velocities are due to the per-step time-integration error, the per-step Lorentz-force-acceleration error, and the per-step collision error.  In the worst-case scenario, these local errors accumulate in a manner such that the global error is one degree lower:
\begin{align*}
p_\alpha = \min\{2,\,p_\text{acc}-1,\,p_\text{coll}-1\} = \min\biggl\{2,\,\frac{\reviewerTwo{s}-3}{2},\,\frac{r-1}{2},\,p_\mathbf{x}\biggr\}.
\end{align*}
Therefore, for second-order accuracy ($p_\alpha=2$), we require $\reviewerTwo{s}=7$ and $r=5$.
If there is no Lorentz-force acceleration, $p_\alpha=\min\{2,\,(\reviewerTwo{s}-3)/2,\,(r-1)/2\}$. Alternatively, if there are no collisions, $p_\alpha=\min\{2,\,(\reviewerTwo{s}-1)/2,\,p_\mathbf{x}\}$, and we require $\reviewerTwo{s}=5$ for $p_\alpha=2$.

Additionally, we are interested in the discrete error norms at $t=T$ for $\phi$:
\begin{align*}
\varepsilon_2^{\phi} = \sqrt{\frac{1}{N_\text{node}} \sum_{q=1}^{N_\text{node}}\Bigl(e_{\phi_q}^{N_{\Delta t}}\Bigr)^2}=\mathcal{O}\bigl(h^{p_\phi}\bigr), 
\qquad
\varepsilon_\infty^{\phi} = \max_q \bigl|e_{\phi_q}^{N_{\Delta t}}\bigr|=\mathcal{O}\bigl(h^{p_\phi}\bigr), 
\end{align*}
where $e_{\phi_q} = \phi_q - \phi_q^M$ is the error in $\phi$ evaluated at node $q$.  From~\eqref{eq:p_phi}, $p_\phi=\min\{2,\,\reviewerTwo{s}/2,\,p_\mathbf{x}\}$.  For $p_\phi=2$, we require $p_\mathbf{x}=2$, which requires $\reviewerTwo{s}=7$ with collisions and $\reviewerTwo{s}=5$ without collisions.  If $\phi$ is decoupled from the particles, $p_\phi=2$.

\subsection{Additional Error Metrics for Collisions} 
\label{sec:additional}

The approaches and analysis in Sections~\ref{sec:coll_src} and~\ref{sec:collisions} detect coding errors that cause the expected outcome from the collision algorithm to not match the analytic source term.  For such coding errors, the expected collision error $\mathbf{e}_{\text{coll}_p}^n$~\eqref{eq:e_coll} is not zero.  However, coding errors that do not modify the expected outcome from the collision algorithm are not detected by this approach.  

Accounting for~\eqref{eq:fm}, \eqref{eq:fv}, \eqref{eq:delta_vp}, and~\eqref{eq:p_unity}, \eqref{eq:vel_change_1a} can be written as
\begin{align}
\bigl\langle\Delta \mathbf{v}_p^M\bigr\rangle_\text{coll}
&{}= \displaystyle\bigl(N_p^\text{cell}-1\bigr) \int_{-\infty}^\infty P_\text{coll}(g)f_\mathbf{v}(\mathbf{v}_q,t) \mathbf{J}(\mathbf{v}_p,\mathbf{v}_q) d\mathbf{v}_q,
\label{eq:vel_change_1f}
\end{align}
where
\begin{align}
\mathbf{J}(\mathbf{v}_p,\mathbf{v}_q) = 
\int_0^{2\pi} \int_0^\pi \Delta \mathbf{v}_p  p(\chi,\epsilon) d\chi d\epsilon = 
\frac{\mathbf{v}_q-\mathbf{v}_p}{2} + \frac{g}{2}\int_0^{2\pi} \int_0^\pi \mathbf{n} p(\chi,\epsilon) d\chi d\epsilon.
\label{eq:J}
\end{align}
From~\eqref{eq:int_np} and~\eqref{eq:constraint_1}, \eqref{eq:J} simplifies to
\begin{align}
\mathbf{J}(\mathbf{v}_p,\mathbf{v}_q) = 
\frac{\mathbf{v}_q-\mathbf{v}_p}{2} .
\label{eq:J2}
\end{align}
If a coding error satisfies~\eqref{eq:J2}, it will not be detected.  A necessary, but not sufficient, condition for~\eqref{eq:J2} to be satisfied is
\begin{align*}
\int_0^{2\pi} \int_0^\pi \mathbf{n}p(\chi,\epsilon) d\chi d\epsilon=\mathbf{0}.
\end{align*}

Therefore, in addition to measuring the convergence of the particle positions and velocities, we also measure the convergence of $p_\chi(\chi)$ and $p_\epsilon(\epsilon)$.  For every collision during a simulation, we compute and record $\chi$ and $\epsilon$ from the post-collision relative velocity $\mathbf{g}'=\mathbf{v}_p'-\mathbf{v}_q'=g\mathbf{n}$: 
\begin{align}
\chi = \cos^{-1}  \frac{g_z'}{g}, \qquad \epsilon = \tan^{-1} \frac{g_y'}{g_x'},
\label{eq:angle_recovery}
\end{align}%
where $\epsilon\in[0,\,2\pi)$ is numerically evaluated from a two-argument arctangent function. 

From these $\chi$ and $\epsilon$ values, we compute
\begin{align*}
\varepsilon_2^{\alpha} = \sqrt{\frac{1}{N_\text{coll}} \sum_{r=1}^{N_\text{coll}}e_{\alpha_r}^2}=\mathcal{O}\bigl(N_\text{coll}^{-1/2}\bigr), 
\qquad
\varepsilon_\infty^{\alpha} = \max_r \bigl|e_{\alpha_r}\bigr|=\mathcal{O}\bigl(N_\text{coll}^{-1/2}\bigr), 
\end{align*}
where $\alpha=\{\chi,\,\epsilon\}$, $N_\text{coll}$ is the total number of collisions during the simulation, and $e_{\alpha_r} = F_{p_\alpha}(\alpha_r) - F_{p_\alpha}^M(\alpha_r)$ is the error between the empirical and analytic cumulative distribution functions.  Letting $\theta$ denote the Heaviside step function, the empirical cumulative distribution function is computed according to
\begin{align*}
F_{p_\alpha}(\alpha) = \frac{1}{N_\text{coll}} \sum_{r=1}^{N_\text{coll}} \theta(\alpha-\alpha_r),
\end{align*}
whereas $F_{p_\alpha}^M$ is the analytic cumulative distribution function evaluated from~\eqref{eq:Fp_eps} and~\eqref{eq:Fp_chi}.  Additionally, when we obtain $\chi$ from~\eqref{eq:angle_recovery}, we use the pre-collision relative speed $g$ so that we confirm $g'=g$ when measuring the convergence of $p_\chi(\chi)$.

\section{Numerical Examples} 
\label{sec:results}

In this section, we demonstrate the approaches presented in Section~\ref{sec:mms} and the convergence rates derived in Section~\ref{sec:error}.

\newlength\thirdwidth \setlength\thirdwidth{\dimexpr\textwidth/3\relax}

\subsection{Manufactured Cross Section and Anisotropy} 

\begin{figure}
\centering
\begin{subfigure}[b]{.5\textwidth}
\raggedright
\includegraphics[scale=.64,clip=true,trim=2.28in 0in 2.842in 0in]{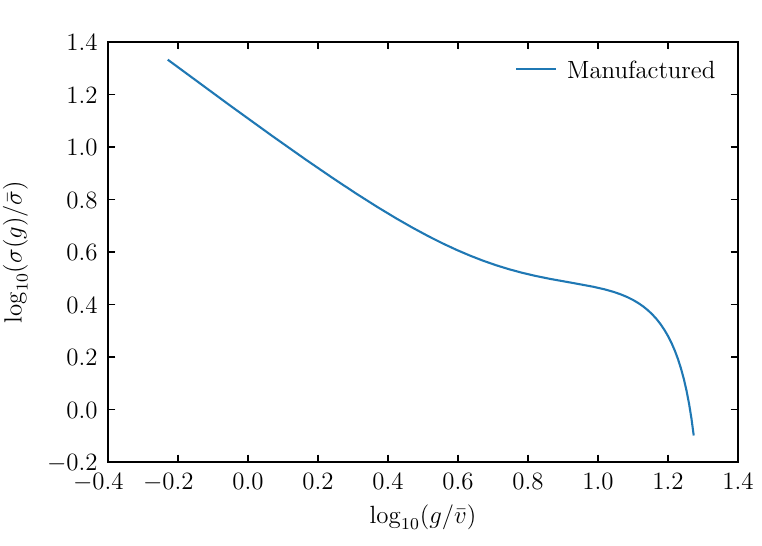}
\caption{Manufactured cross section\vpad}
\label{fig:cross_section}
\end{subfigure}\hfill
\begin{subfigure}[b]{.5\textwidth}
\raggedleft
\includegraphics[scale=.64,clip=true,trim=2.28in 0in 2.842in 0in]{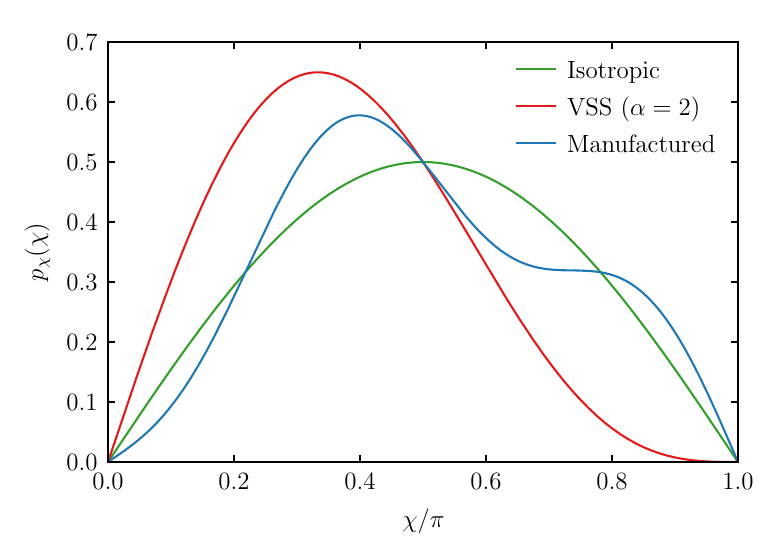}
\caption{Manufactured anisotropy\vpad}
\label{fig:anisotropy}
\end{subfigure}
\caption{Manufactured cross section and anisotropy.}
\vskip-\dp\strutbox
\label{fig:cross_section_anisotropy}
\end{figure}

The manufactured cross section is constructed with $N_\sigma=3$ and is shown in Figure~\ref{fig:cross_section}, where $\bar{\sigma}=1$~\AA$^2$ and $\bar{v}=10^6$~m/s.


For the manufactured anisotropy, we consider the variable soft sphere model~\cite{koura_1991,koura_1992}, where the polar angle is given by
\begin{align*}
\chi = 2\cos^{-1} \biggl(\frac{b}{d}\biggr)^{1/\alpha}, 
\end{align*} 
with $b/d=\sqrt{\xi_\chi}$ and $1\le\alpha\le 2$.  Consequently,
\begin{align}
\chi = F_{p_\chi}^{-1}(\xi_\chi)=2\cos^{-1} \xi_\chi^{1/(2\alpha)}, \qquad
F_{p_\chi}(\chi)= 1 - \cos^{2\alpha}\frac{\chi}{2}, \qquad 
p_\chi(\chi)=\alpha\cos^{2\alpha-1}\frac{\chi}{2}\sin\frac{\chi}{2}. 
\label{eq:vss}
\end{align}
When $\alpha=1$, the variable soft sphere model simplifies to the isotropic variable hard sphere model~\cite{bird_1994}; when $\alpha=2$, the anisotropy is maximized.  

In~\eqref{eq:anisotropy_opt}, we set $\bar{p}_\chi(\chi)$ to that of the variable soft sphere model~\eqref{eq:vss} when $\alpha=2$:
\begin{align}
\bar{p}_\chi(\chi)=2\cos^{3}\frac{\chi}{2}\sin\frac{\chi}{2}.
\label{eq:pbar}
\end{align}
With $\bar{p}_\chi(\chi)$~\eqref{eq:pbar}, \eqref{eq:anisotropy_opt} is minimized with $C_0=1/2$ and $C_1=12/29$ in~\eqref{eq:pchi_ms}, and
\begin{align*}
p_\chi(\chi) = \frac{1}{29}\biggl(\frac{29}{2} + 12\cos\chi  -20\cos^3\chi\biggr)\sin\chi, \qquad
F_{p_\chi}(\chi) = \sin^2\frac{\chi}{2} - \frac{1}{58}\bigl(3+5\cos 2\chi\bigr)\sin^2\chi.
\end{align*}
Figure~\ref{fig:anisotropy} shows $p_\chi(\chi)$ for isotropic, variable-soft-sphere anisotropic ($\alpha=2$), and the manufactured anisotropic scattering.
To compute the $F_{p_\chi}^{-1}$, we set $u=\sin^2 (\chi/2)$, such that
\begin{align*}
F_{p_\chi}(u) = \frac{1}{29}(13u + 96 u^2 - 160 u^3 + 80 u^4),
\end{align*}
and
\begin{align*}
u(\xi_\chi) = \frac{1}{2}\Biggl(1-\sqrt{b(\xi_\chi)}+\sqrt{\frac{3}{5}-b\bigl(\xi_\chi\bigr)+\frac{29}{40\sqrt{b(\xi_\chi)}}}\Biggr),
\end{align*}
where
\begin{align*}
a(\xi_\chi) &{}=\frac{1}{40\cdot 3^{2/3}}\Bigl(81909-83520\xi_\chi + \sqrt{298924665+55680\xi_\chi\bigl[376377+2320\xi_\chi\bigl(-429+290\xi_\chi\bigr)\bigr]}\Bigr)^{1/3},
\\[.5em]
b(\xi_\chi) &{}= \frac{1}{5} + a(\xi_\chi) + \frac{161-290 \xi_\chi}{600 a(\xi_\chi)},
\end{align*}
and
\begin{align}
\chi = F_{p_\chi}^{-1}(\xi_\chi) = 2\sin^{-1}\sqrt{u(\xi_\chi)}.
\label{eq:fpiv_ms}
\end{align}

\begin{figure}[!t]
\centering
\begin{subfigure}[b]{\thirdwidth}
\raggedright
\includegraphics[scale=.64,clip=true,trim=2.28in 0in 2.842in 0in]{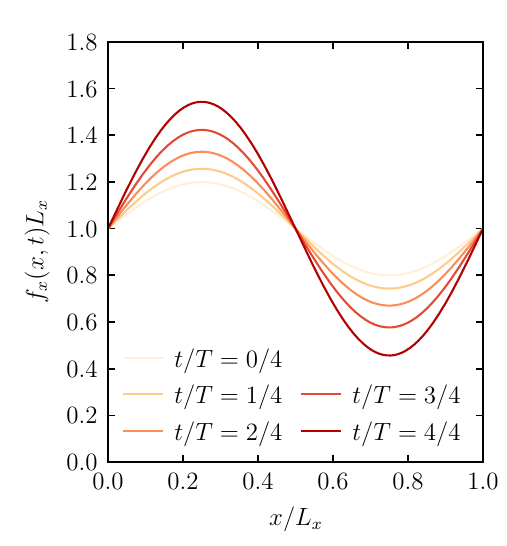}
\caption{$f_x(x,t)$\vpad}
\label{fig:f_x_history}
\end{subfigure}\hfill
\begin{subfigure}[b]{\thirdwidth}
\centering
\includegraphics[scale=.64,clip=true,trim=2.28in 0in 2.842in 0in]{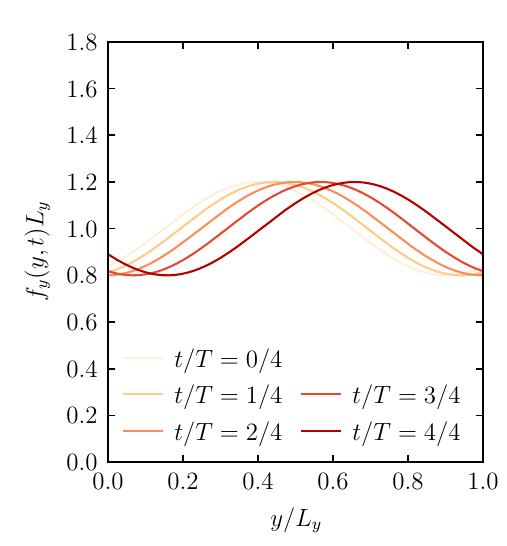}
\caption{$f_y(y,t)$\vpad}
\label{fig:f_y_history}
\end{subfigure}\hfill
\begin{subfigure}[b]{\thirdwidth}
\raggedleft
\includegraphics[scale=.64,clip=true,trim=2.28in 0in 2.842in 0in]{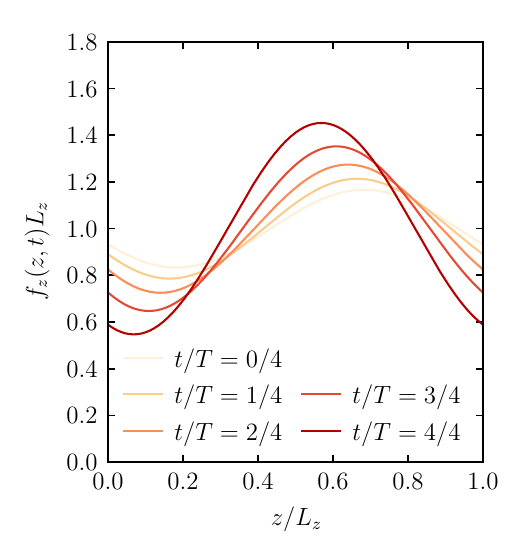}
\caption{$f_z(z,t)$\vpad}
\label{fig:f_z_history}
\end{subfigure}
\\
\begin{subfigure}[b]{\thirdwidth}
\raggedright
\includegraphics[scale=.64,clip=true,trim=2.28in 0in 2.842in 0in]{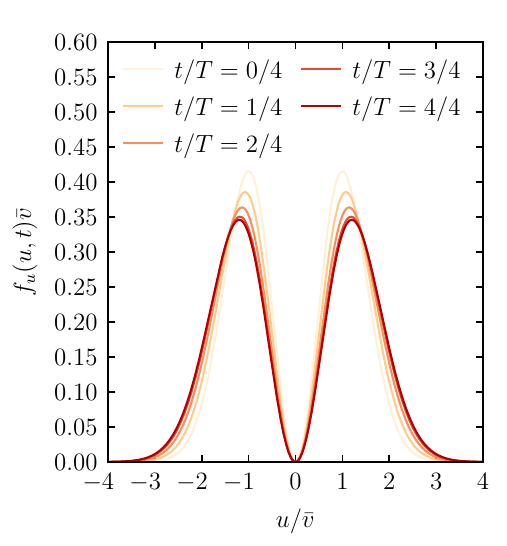}
\caption{$f_u(u,t)$\vpad}
\label{fig:f_u_history}
\end{subfigure}\hfill
\begin{subfigure}[b]{\thirdwidth}
\centering
\includegraphics[scale=.64,clip=true,trim=2.28in 0in 2.842in 0in]{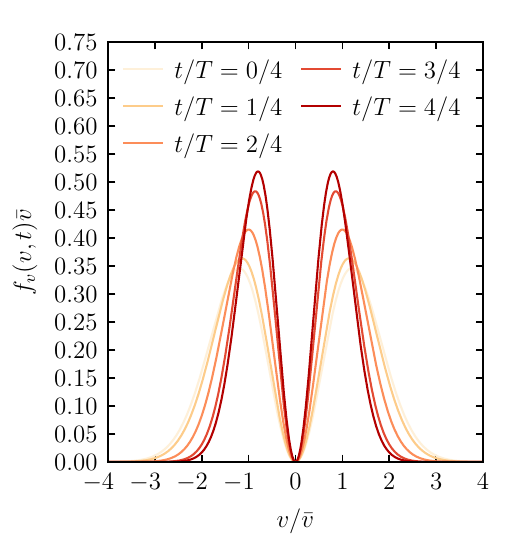}
\caption{$f_v(v,t)$\vpad}
\label{fig:f_v_history}
\end{subfigure}\hfill
\begin{subfigure}[b]{\thirdwidth}
\raggedleft
\includegraphics[scale=.64,clip=true,trim=2.28in 0in 2.842in 0in]{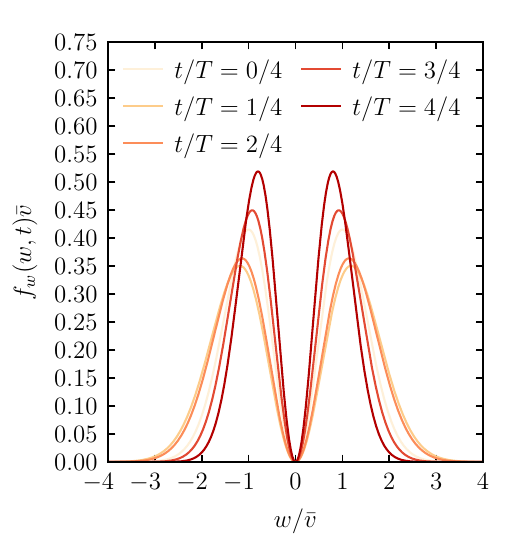}
\caption{$f_w(w,t)$\vpad}
\label{fig:f_w_history}
\end{subfigure}
\caption{The manufactured particle distribution function $f^M(\mathbf{x},\mathbf{v},t) = f_\mathbf{x}(\mathbf{x},t) f_\mathbf{v}(\mathbf{v},t)$ at multiple times.}
\vskip-\dp\strutbox
\label{fig:fm_history}
\end{figure}

\begin{figure}[!t]
\centering
\begin{subfigure}[b]{\thirdwidth}
\raggedright
\includegraphics[scale=.64,clip=true,trim=2.28in 0in 2.842in 0in]{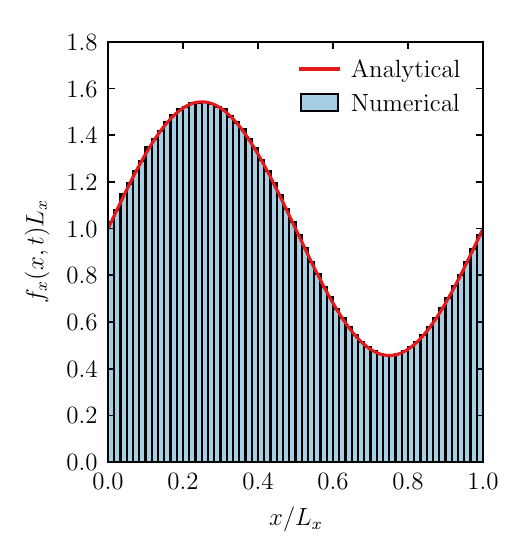}
\caption{$f_x(x,t)$\vpad}
\label{fig:f_x}
\end{subfigure}\hfill
\begin{subfigure}[b]{\thirdwidth}
\centering
\includegraphics[scale=.64,clip=true,trim=2.28in 0in 2.842in 0in]{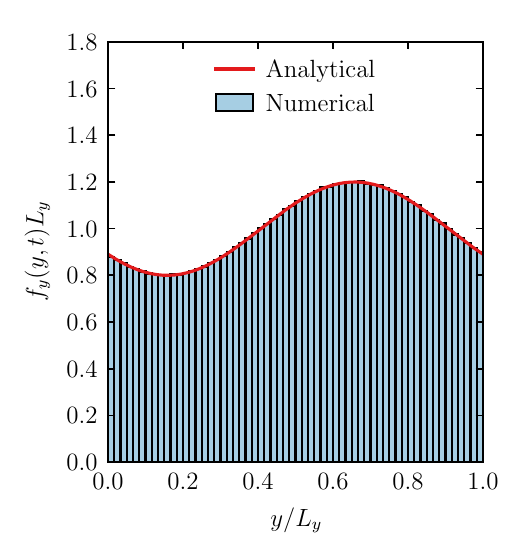}
\caption{$f_y(y,t)$\vpad}
\label{fig:f_y}
\end{subfigure}\hfill
\begin{subfigure}[b]{\thirdwidth}
\raggedleft
\includegraphics[scale=.64,clip=true,trim=2.28in 0in 2.842in 0in]{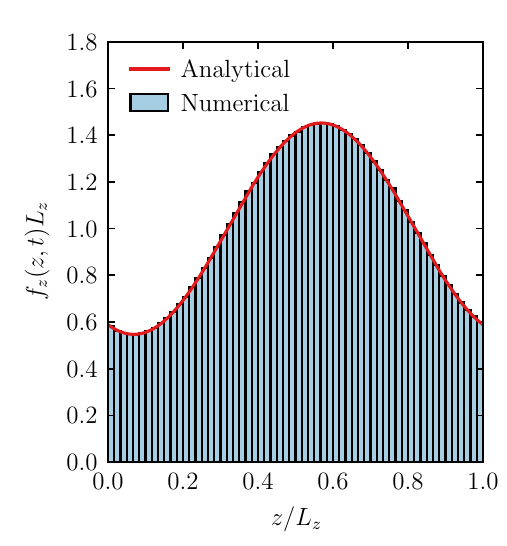}
\caption{$f_z(z,t)$\vpad}
\label{fig:f_z}
\end{subfigure}
\\
\begin{subfigure}[b]{\thirdwidth}
\raggedright
\includegraphics[scale=.64,clip=true,trim=2.28in 0in 2.842in 0in]{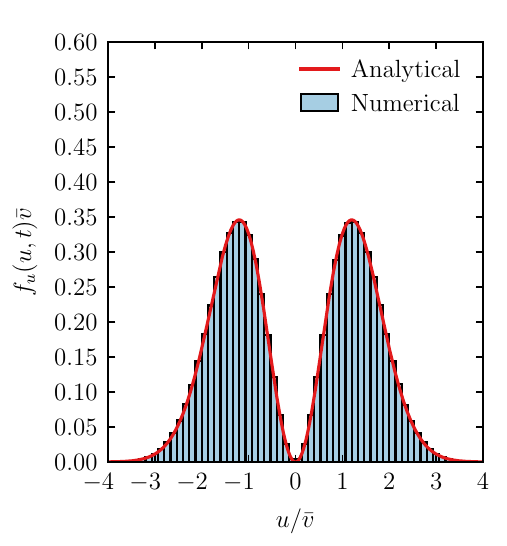}
\caption{$f_u(u,t)$\vpad}
\label{fig:f_u}
\end{subfigure}\hfill
\begin{subfigure}[b]{\thirdwidth}
\centering
\includegraphics[scale=.64,clip=true,trim=2.28in 0in 2.842in 0in]{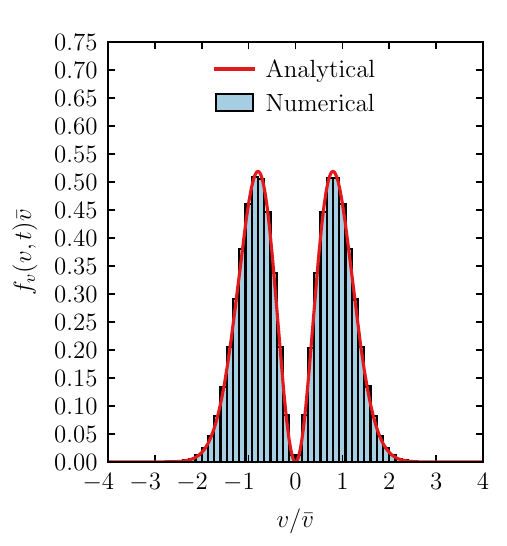}
\caption{$f_v(v,t)$\vpad}
\label{fig:f_v}
\end{subfigure}\hfill
\begin{subfigure}[b]{\thirdwidth}
\raggedleft
\includegraphics[scale=.64,clip=true,trim=2.28in 0in 2.842in 0in]{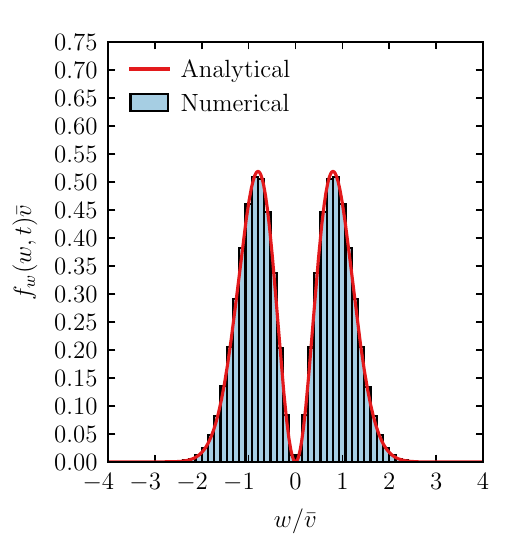}
\caption{$f_w(w,t)$\vpad}
\label{fig:f_w}
\end{subfigure}
\caption{The manufactured particle distribution function $f^M(\mathbf{x},\mathbf{v},t)$ and the distribution of computational particles for the finest discretization at $t=T$ for the fully coupled collisional case.}
\vskip-\dp\strutbox
\label{fig:fm}
\end{figure}

\subsection{Manufactured Particle Distribution Function and Electric Potential}

\newcommand{\WidestEntryA}{$\displaystyle\frac{e^{t/T}}{5}$}%
\newcommand{\SetToWidestA}[1]{\makebox[\widthof{\WidestEntryA}]{$#1$}}%
\newcommand{\WidestEntryB}{$\displaystyle\frac{3e^{t/T}}{20}$}%
\newcommand{\SetToWidestB}[1]{\makebox[\widthof{\WidestEntryB}]{$#1$}}%

As described in Section~\ref{sec:mms}, the manufactured distribution function takes the form of~\eqref{eq:fm}.  
For $f_{\mathbf{v}}(\mathbf{v},t)$~\eqref{eq:fv}, we set
\begin{align*}
\hat{u}(t) = \biggl(1+\frac{1}{5}\sin\frac{  \pi t}{2 T}\biggr)\bar{v}, \qquad
\hat{v}(t) = \biggl(1+\frac{1}{5}\cos\frac{  \pi t}{  T}\biggr)\bar{v}, \qquad
\hat{w}(t) = \biggl(1+\frac{1}{5}\sin\frac{3 \pi t}{2 T}\biggr)\bar{v}.
\end{align*}
For $f_\mathbf{x}(\mathbf{x},t)$~\eqref{eq:fx}, we consider a periodic domain with
\begin{align*}
f_x(x,t) &{}= \frac{1}{L_x}\biggl(1 + \frac{e^{t/T}}{5}\sin\biggl(\frac{2\pi x}{L_x}\biggr)\biggr), \\
f_y(y,t) &{}= \frac{1}{L_y}\biggl(1 + \SetToWidestA{\displaystyle\frac{1}{5}}\sin\biggl(2\pi\biggl[\frac{y}{L_y} - \frac{3e^{t/T}}{20} \biggr]\biggr)\biggr), \\
f_z(z,t) &{}= \frac{1}{L_z}\biggl(1 - \frac{e^{t/T}}{6}\sin\biggl(2\pi\biggl[\frac{z}{L_z} + \SetToWidestB{\displaystyle\frac{e^{t/T}}{15}}\biggr]\biggr)\biggr).
\end{align*}
We additionally set $N=10^{20}$, $L_{x_i}=L=3/2$~m, $T=L/(10\bar{v})$, $q=e$, and $m=3\times 10^8 m_e$, where $m_e$ \rereading{is the mass of an electron and $e$ is the elementary charge}.  For $(\sigma g)_\text{max}$ in the collision algorithm, we compute the maximum of $\sigma(g) g$ over $g\in[0,\,g_\text{max}]$, where $g_\text{max}=10\sqrt{3}\bar{v}$. Figure~\ref{fig:fm_history} shows the manufactured distribution function dependencies at multiple times.

For the electric potential, we manufacture
\begin{align*}
\phi^M(\mathbf{x},t) = \bar{\phi}e^{t/(2T)}
\sin\biggl(2\pi\biggl[\frac{x}{L_x}-\frac{1}{7}\biggr]\biggr)
\sin\biggl(2\pi\biggl[\frac{y}{L_y}-\frac{1}{5}\biggr]\biggr)
\sin\biggl(2\pi\biggl[\frac{z}{L_z}-\frac{1}{3}\biggr]\biggr), 
\end{align*}
where $\bar{\phi}=10^{10}$~V.

\begin{table}[!t]
\centering
\begin{tabular}{c c c c c c c c c c}
\toprule 
      &              &                     &                       & \multicolumn{4}{c}{Collisional}                             & \multicolumn{2}{c}{Collisionless}     \\
                                                                     \cmidrule(r){5-8}                                                     \cmidrule(l){9-10}
Disc. & $T/\Delta t$ & $N_{\text{cell}_i}$ & $N_\text{cell}$       & $N_\text{avg}$ & $1/P_{\text{coll}_\text{max}}$ & $N_p$              & $N_p/N_\text{cell}$ & $N_p$           & $N_p/N_\text{cell}$ \\ \midrule
1     & \pz8         & \pz8                & \pz\pz\pc512          & \pz\pc\pz32    &   \pz\pz\pc108.28              & \pz\pz\pc\pz10,240 & \pz\pz20.00         & \pz\pc\pz10,240 & \pz20.00            \\
2     &   12         &   12                &     \pz1,728          &   \pz\pc243    &   \pz\pz\pc822.27              &   \pz\pz\pc174,960 &   \pz101.25         & \pz\pc\pz77,760 & \pz45.00            \\
3     &   16         &   16                &     \pz4,096          &     1,024      &     \pz3,465.04                &       \pz1,310,720 &   \pz320.00         &   \pz\pc327,680 & \pz80.00            \\
4     &   20         &   20                &     \pz8,000          &     3,125      &       10,574.47                &       \pz6,250,000 &   \pz781.25         &       1,000,000 &   125.00            \\
5     &   24         &   24                &       13,824          &     7,776      &       26,312.66                &         22,394,880 &     1620.00         &       2,488,320 &   180.00            \\ \bottomrule
\end{tabular}
\caption{Discretizations.}
\label{tab:discretizations}
\end{table}

\subsection{Discretizations} 

We consider cases with and without collisions with varying degrees of coupling between the particles and field over a series of five discretizations described in Table~\ref{tab:discretizations}.  From Section~\ref{sec:error_accumulation}, for second-order accuracy ($p=2$), in~\eqref{eq:refinements}, we set $\reviewerTwo{s}=7$ and $r=5$ for the collisional case and $\reviewerTwo{s}=5$ for the collisionless case.  For the collisional case, Table~\ref{tab:discretizations} shows that~\eqref{eq:navg_ineq} is satisfied. 

\reviewerOne{The collisional cases are substantially more computationally expensive than the collisionless cases.  They require considerably more particles ($N_p\sim h^{-7}$) than the collisionless case ($N_p\sim h^{-5}$), while also requiring the collision algorithm to be run $N_\text{avg}\sim h^{-5}$ times for each evaluation.  Nonetheless, even the collisionless case is expensive because $N_p\sim h^{-5}$ grows rapidly with refinement.  Therefore, we increase the number of time steps and the one-dimensional mesh size incrementally rather than by a constant refinement factor.}

At $t=T$, Figure~\ref{fig:fm} shows how the distribution of the computational particles compares with the manufactured distribution function for the fully coupled collisional case.

\begin{figure}
\centering
\begin{subfigure}[b]{.5\textwidth}
\raggedright
\includegraphics[scale=.64,clip=true,trim=2.28in 0in 2.842in 0in]{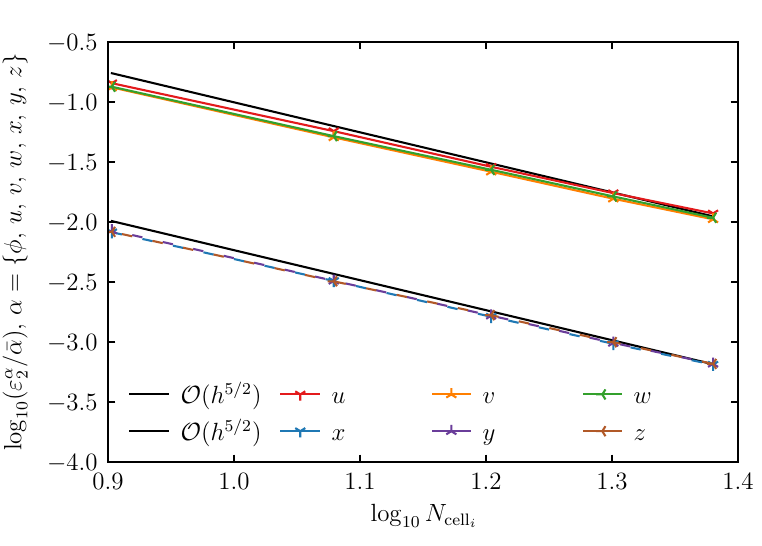}
\caption{Reference, $\varepsilon_2$\vpad}
\label{fig:just_coll_l2}
\end{subfigure}\hfill
\begin{subfigure}[b]{.5\textwidth}
\raggedleft
\includegraphics[scale=.64,clip=true,trim=2.28in 0in 2.842in 0in]{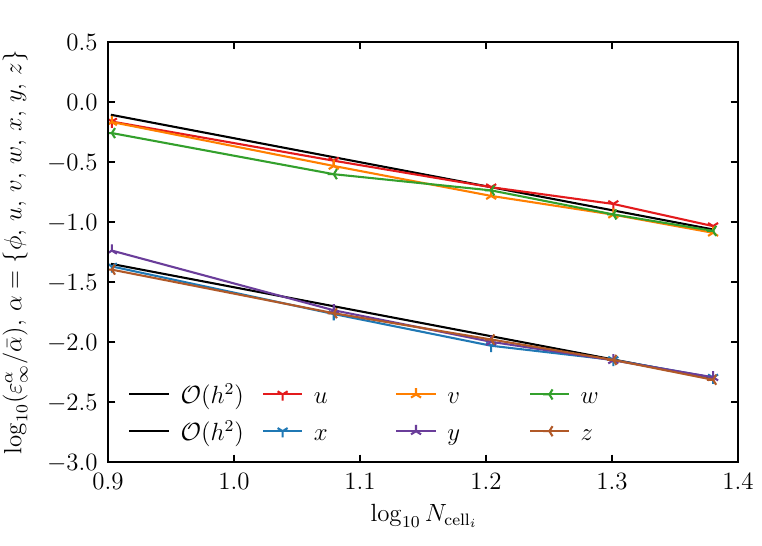}
\caption{Reference, $\varepsilon_\infty$\vpad}
\label{fig:just_coll_linf}
\end{subfigure}
\\
\begin{subfigure}[b]{.5\textwidth}
\raggedright
\includegraphics[scale=.64,clip=true,trim=2.28in 0in 2.842in 0in]{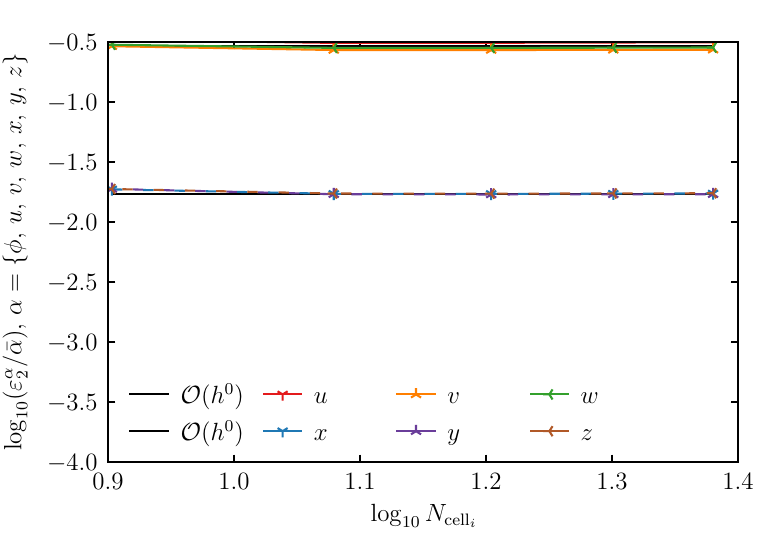}
\caption{Error 1, $\varepsilon_2$\vpad}
\label{fig:code_bug_1_l2}
\end{subfigure}\hfill
\begin{subfigure}[b]{.5\textwidth}
\raggedleft
\includegraphics[scale=.64,clip=true,trim=2.28in 0in 2.842in 0in]{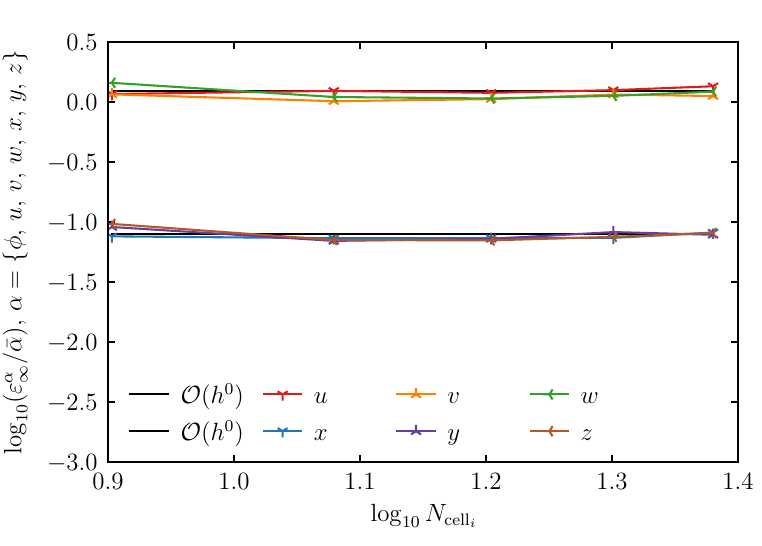}
\caption{Error 1, $\varepsilon_\infty$\vpad}
\label{fig:code_bug_1_linf}
\end{subfigure}
\\
\begin{subfigure}[b]{.5\textwidth}
\raggedright
\includegraphics[scale=.64,clip=true,trim=2.28in 0in 2.842in 0in]{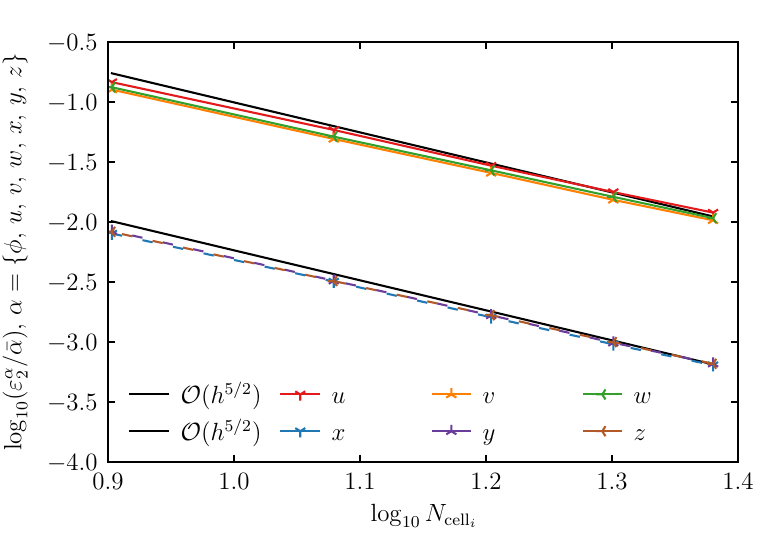}
\caption{Error 2, $\varepsilon_2$\vpad}
\label{fig:code_bug_2_l2}
\end{subfigure}\hfill
\begin{subfigure}[b]{.5\textwidth}
\raggedleft
\includegraphics[scale=.64,clip=true,trim=2.28in 0in 2.842in 0in]{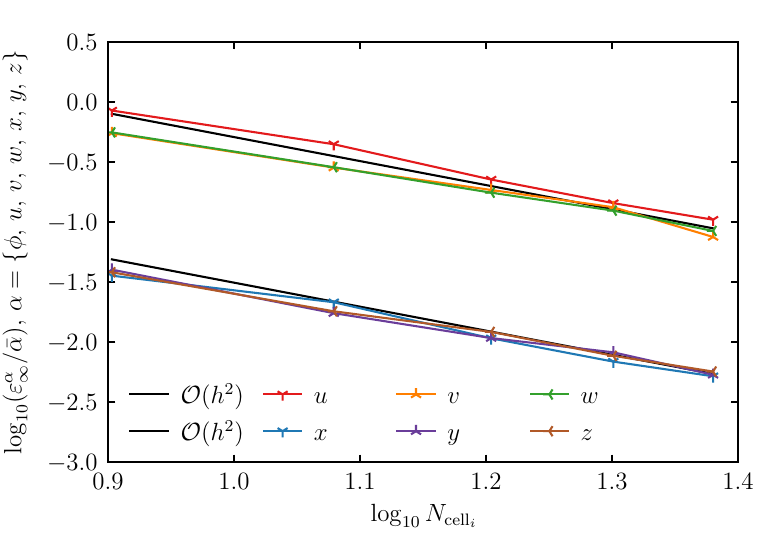}
\caption{Error 2, $\varepsilon_\infty$\vpad}
\label{fig:code_bug_2_linf}
\end{subfigure}
\caption{Collisional: convergence of the isolated collision error at $t=T$ for different norms, with and without coding errors.}
\vskip-\dp\strutbox
\end{figure}

\begin{figure}
\centering
\begin{subfigure}[b]{.5\textwidth}
\raggedright
\includegraphics[scale=.64,clip=true,trim=2.28in 0in 2.842in 0in]{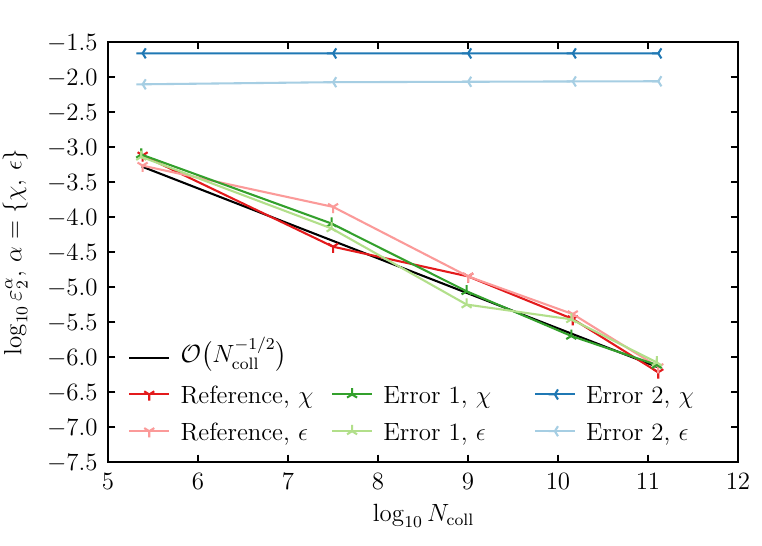}
\caption{Uncoupled, $\varepsilon_2$\vpad}
\label{fig:angles_l2}
\end{subfigure}\hfill
\begin{subfigure}[b]{.5\textwidth}
\raggedleft
\includegraphics[scale=.64,clip=true,trim=2.28in 0in 2.842in 0in]{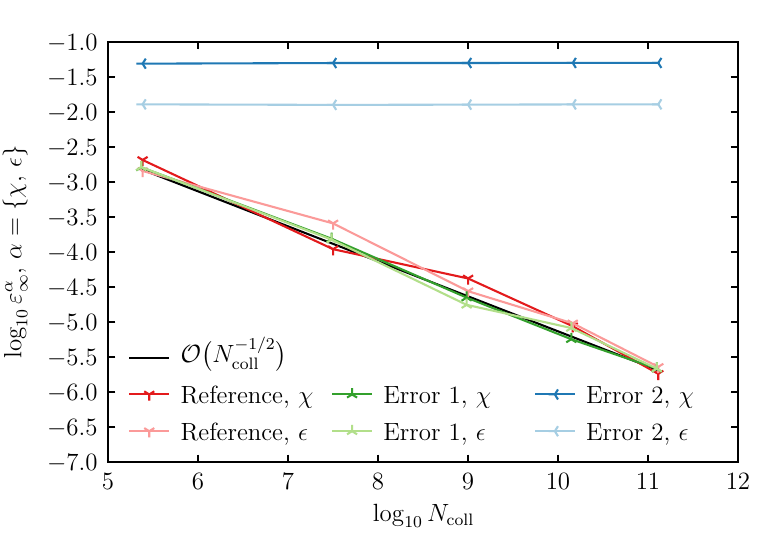}
\caption{Uncoupled, $\varepsilon_\infty$\vpad}
\label{fig:angles_linf}
\end{subfigure}
\caption{Collisional: convergence of the scattering-angle distributions for different norms, with and without coding errors.}
\vskip-\dp\strutbox
\end{figure}

\begin{figure}
\centering
\begin{subfigure}[b]{.5\textwidth}
\raggedright
\includegraphics[scale=.64,clip=true,trim=2.28in 0in 2.842in 0in]{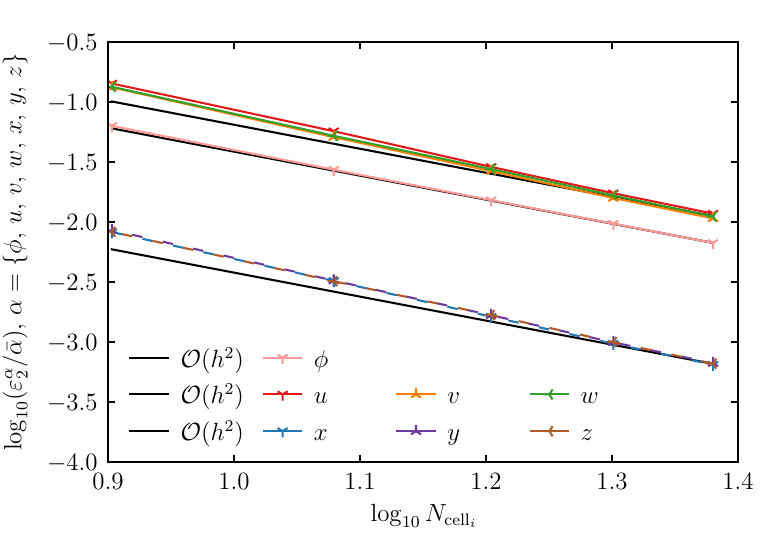}
\caption{Uncoupled, $\varepsilon_2$\vpad}
\label{fig:uncoupled_coll_l2}
\end{subfigure}\hfill
\begin{subfigure}[b]{.5\textwidth}
\raggedleft
\includegraphics[scale=.64,clip=true,trim=2.28in 0in 2.842in 0in]{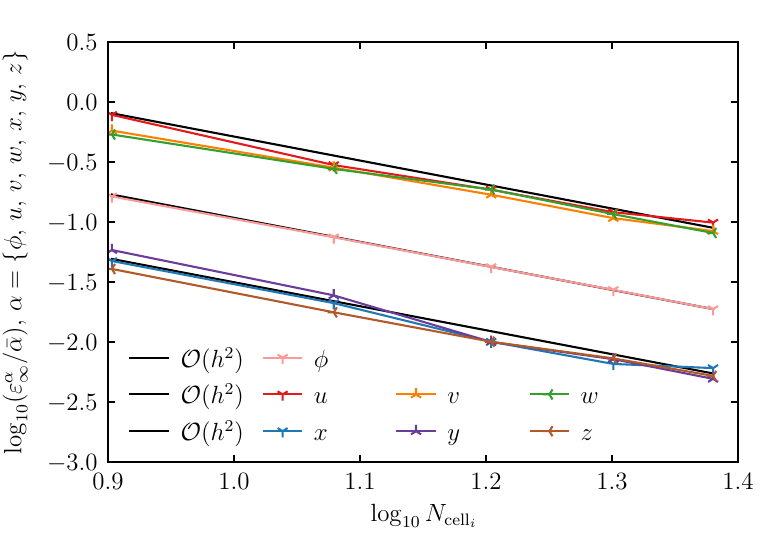}
\caption{Uncoupled, $\varepsilon_\infty$\vpad}
\label{fig:uncoupled_coll_linf}
\end{subfigure}
\\
\begin{subfigure}[b]{.5\textwidth}
\raggedright
\includegraphics[scale=.64,clip=true,trim=2.28in 0in 2.842in 0in]{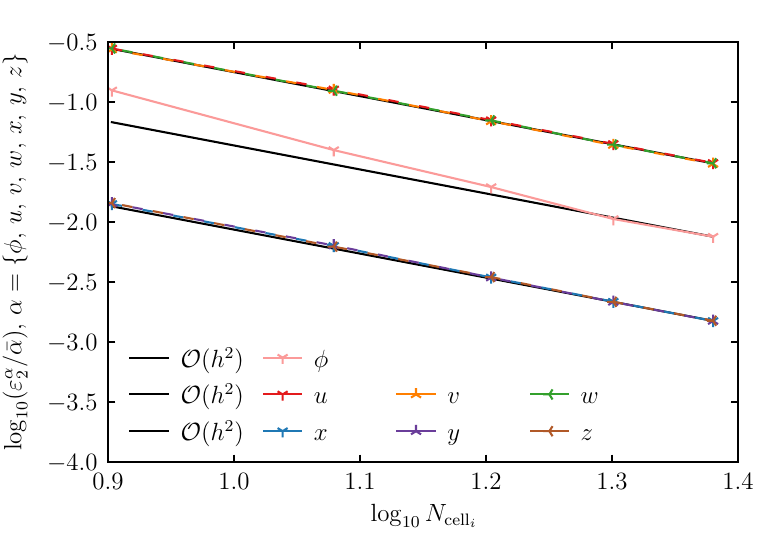}
\caption{One-way coupled, $\varepsilon_2$\vpad}
\label{fig:oneway_coll_l2}
\end{subfigure}\hfill
\begin{subfigure}[b]{.5\textwidth}
\raggedleft
\includegraphics[scale=.64,clip=true,trim=2.28in 0in 2.842in 0in]{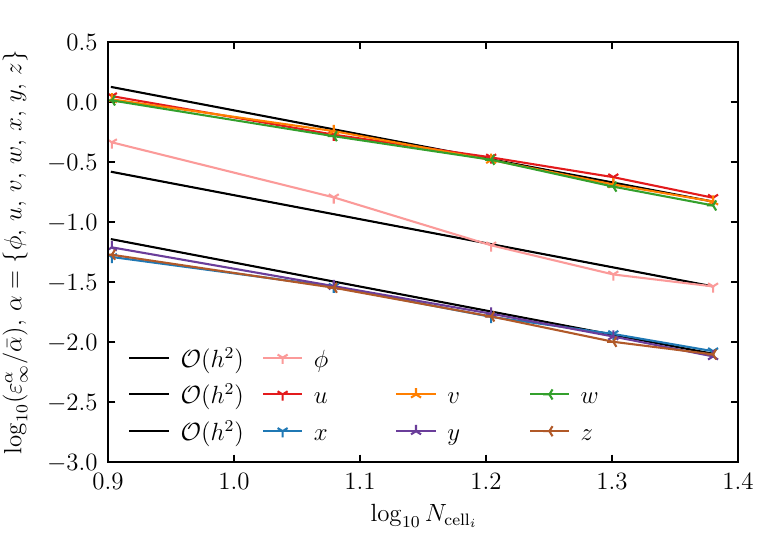}
\caption{One-way coupled, $\varepsilon_\infty$\vpad}
\label{fig:oneway_coll_linf}
\end{subfigure}
\\
\begin{subfigure}[b]{.5\textwidth}
\raggedright
\includegraphics[scale=.64,clip=true,trim=2.28in 0in 2.842in 0in]{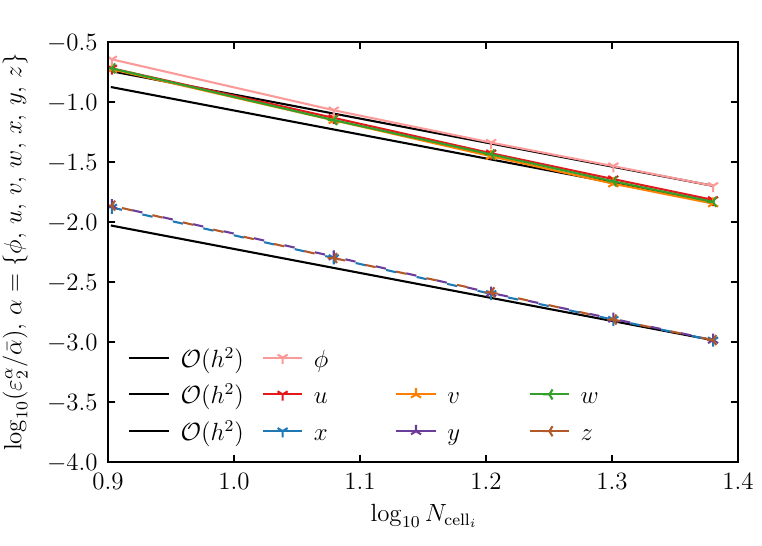}
\caption{Fully coupled, $\varepsilon_2$\vpad}
\label{fig:coupled_coll_l2}
\end{subfigure}\hfill
\begin{subfigure}[b]{.5\textwidth}
\includegraphics[scale=.64,clip=true,trim=2.28in 0in 2.842in 0in]{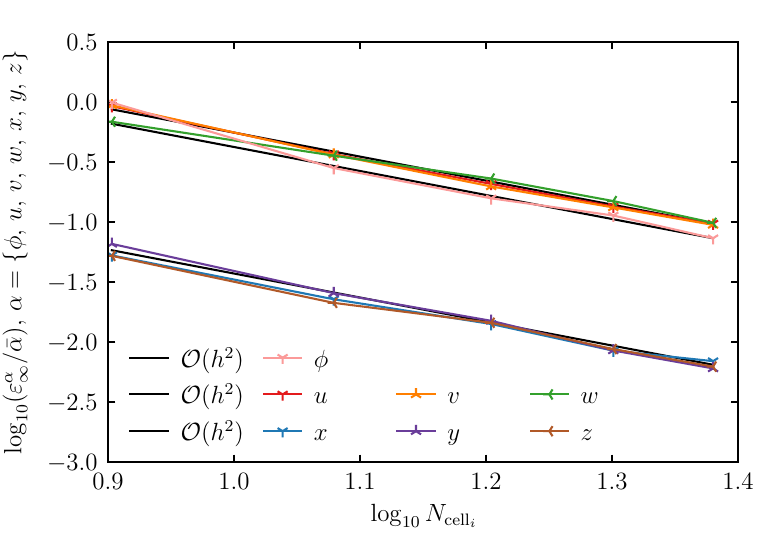}
\caption{Fully coupled, $\varepsilon_\infty$\vpad}
\label{fig:coupled_coll_linf}
\end{subfigure}
\caption{Collisional: convergence of the error at $t=T$ for different norms.}
\vskip-\dp\strutbox
\label{fig:error_coll}
\end{figure}

\begin{figure}[!t]
\centering
\begin{subfigure}[b]{.5\textwidth}
\raggedright
\includegraphics[scale=.64,clip=true,trim=2.28in 0in 2.842in 0in]{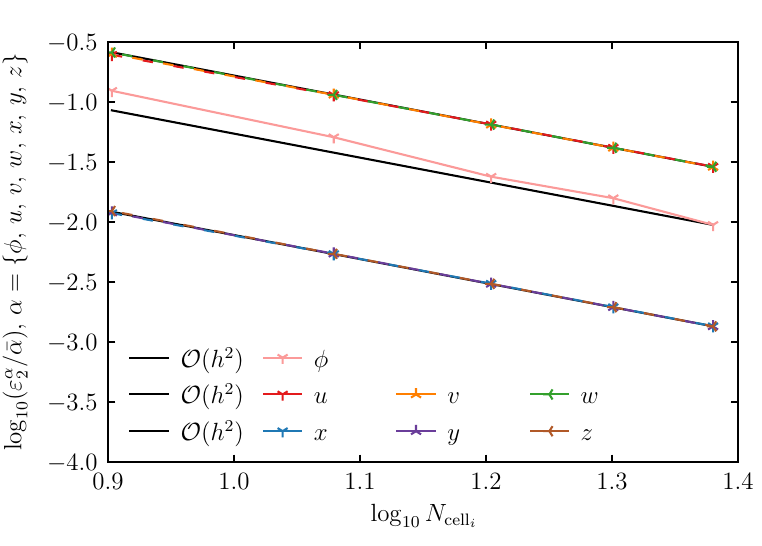}
\caption{One-way coupled, $\varepsilon_2$\vpad}
\label{fig:oneway_nocoll_l2}
\end{subfigure}\hfill
\begin{subfigure}[b]{.5\textwidth}
\raggedleft
\includegraphics[scale=.64,clip=true,trim=2.28in 0in 2.842in 0in]{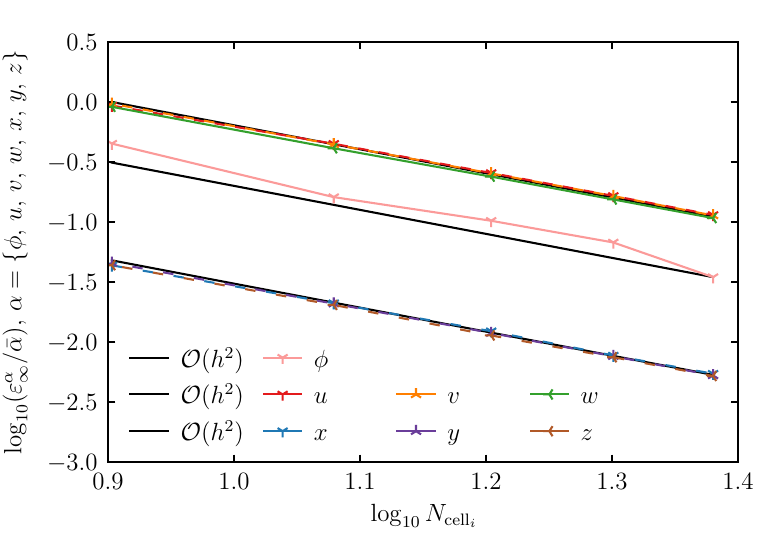}
\caption{One-way coupled, $\varepsilon_\infty$\vpad}
\label{fig:oneway_nocoll_linf}
\end{subfigure}
\\
\begin{subfigure}[b]{.5\textwidth}
\raggedright
\includegraphics[scale=.64,clip=true,trim=2.28in 0in 2.842in 0in]{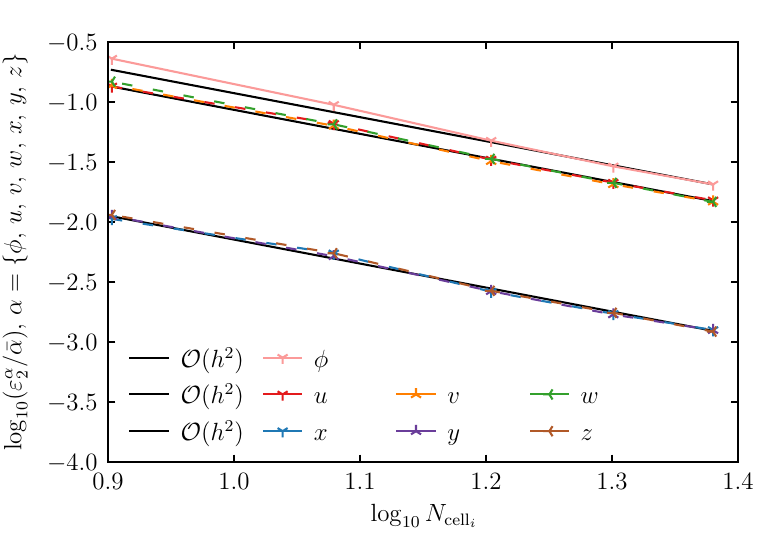}
\caption{Fully coupled, $\varepsilon_2$\vpad}
\label{fig:coupled_nocoll_l2}
\end{subfigure}\hfill
\begin{subfigure}[b]{.5\textwidth}
\includegraphics[scale=.64,clip=true,trim=2.28in 0in 2.842in 0in]{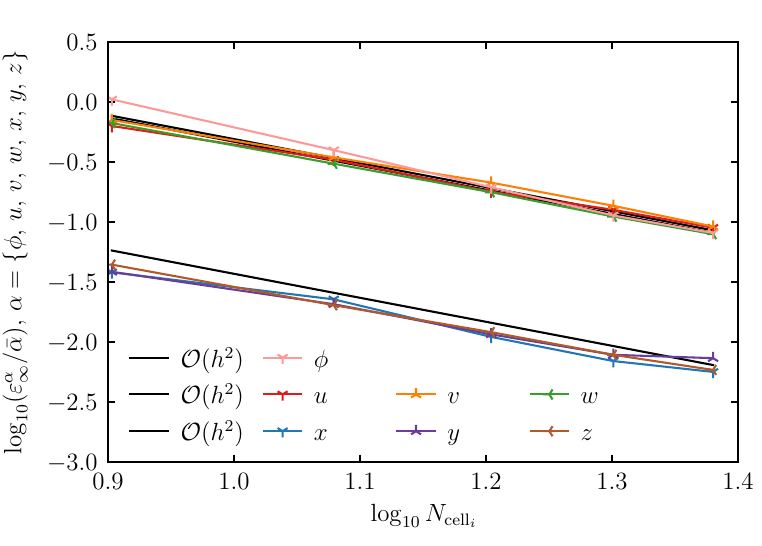}
\caption{Fully coupled, $\varepsilon_\infty$\vpad}
\label{fig:coupled_nocoll_linf}
\end{subfigure}
\caption{Collisionless: convergence of the error at $t=T$ for different norms.}
\vskip-\dp\strutbox
\label{fig:error_nocoll}
\end{figure}

\subsubsection{Collisional} 

We begin by measuring the error convergence for the collisional case.  Given the many sources of error and their interactions, it is insightful to isolate the error sources to some degree.  To that end, we can isolate the particle error due to collisions by setting 
\begin{alignat*}{7}
\frac{1}{2}\Delta t\bigl(&&\dot{\mathbf{v}}_p^M & \bigr)^{n\phantom{{}+0/1}} = \bigl(&\mathbf{v}_p^M&\bigr)^{n+1/2}&&{}-\bigl(&\mathbf{v}_p^M&\bigr)^n, \\
           \Delta t\bigl(&&\dot{\mathbf{x}}_p^M & \bigr)^{n+1/2}             = \bigl(&\mathbf{x}_p^M&\bigr)^{n+1}  &&{}-\bigl(&\mathbf{x}_p^M&\bigr)^n, \\
\frac{1}{2}\Delta t\bigl(&&\dot{\mathbf{v}}_p^M & \bigr)^{n+1\phantom{/1}}   = \bigl(&\mathbf{v}_p^M&\bigr)^{n+1}  &&{}-\bigl(&\mathbf{v}_p^M&\bigr)^{n+1/2}
\end{alignat*}
in~\eqref{eq:velocity_verlet} to eliminate the time-integration error, and we can set $q/m=0$ so that the electric field does not affect the particles.  In Figure~\ref{fig:just_coll_linf}, the collision error is $\mathcal{O}(h^2)$ for $\varepsilon_\infty$, as expected. However, in Figure~\ref{fig:just_coll_l2}, for $\varepsilon_2$, the accumulation of the collision error better resembles that from a random walk ($\mathcal{O}(h^{5/2})$) than from the worst-case scenario ($\mathcal{O}(h^2)$) described in Section~\ref{sec:error_accumulation}.  This reduced accumulation is due to the mean collision error being zero, the collision-algorithm runs being independent, and $f_{\mathbf{v}}(\mathbf{v},t)$ not varying significantly with time. In Figures~\ref{fig:angles_l2} and~\ref{fig:angles_linf}, the distributions of the scattering angles for this reference case converge according to $\mathcal{O}\bigl(N_\text{coll}^{-1/2}\bigr)$, as expected.

To demonstrate the ability of the approaches presented in Sections~\ref{sec:mms} and~\ref{sec:error} to reveal coding errors, we introduce two coding errors.  
For Error~1, in the collision algorithm, we incorrectly compute $\mathbf{v}_{\text{cm}} = (\mathbf{v}_p - \mathbf{v}_q)/2$.  Error~1 modifies the expected outcome from the collision algorithm.  
As a result, Figures~\ref{fig:code_bug_1_l2} and~\ref{fig:code_bug_1_linf} show that the error does not converge, consequently revealing the coding error.
However, Error~1 does not modify the values of $\chi$ or $\epsilon$ as described in Section~\ref{sec:additional}.  Therefore, the scattering-angle distributions converge at the expected rate in Figures~\ref{fig:angles_l2} and~\ref{fig:angles_linf}, such that those metrics do not reveal the coding error.  
For Error~2, we modify the post-collision velocity update so that the velocities are swapped with probability $\beta$ or not updated with probability $(1-\beta)$.  Consequently, \eqref{eq:delta_vp} is effectively replaced by
\begin{align*}
\Delta\mathbf{v}_p = \beta(\mathbf{v}_q-\mathbf{v}_p),
\end{align*}
and~\eqref{eq:J} becomes
\begin{align}
\mathbf{J}(\mathbf{v}_p,\mathbf{v}_q) = 
\int_0^{2\pi} \int_0^\pi \Delta \mathbf{v}_p  p(\chi,\epsilon) d\chi d\epsilon = 
\beta(\mathbf{v}_q-\mathbf{v}_p).
\label{eq:J3}
\end{align}
We set $\beta=1/2$, such that \eqref{eq:J3} is equal to~\eqref{eq:J2} so that Error~2 does not modify the expected outcome from the collision algorithm.  As a result, the error converges in Figures~\ref{fig:code_bug_2_l2} and~\ref{fig:code_bug_2_linf}, and the coding error is undetected by those metrics.  On the other hand, Figures~\ref{fig:angles_l2} and~\ref{fig:angles_linf} show the scattering-angle distributions fail to converge, thus revealing the coding error.  Between these three cases -- the reference case, Error 1, and Error 2 -- only the reference case converges for both metrics.

Restoring the time-integration error, Figures~\ref{fig:uncoupled_coll_l2} and~\ref{fig:uncoupled_coll_linf} show how the particles and potential converge when the particles and field are uncoupled.  By setting $q/m=0$, the electric field does not affect the particles, and by setting $q=0$, the particles do not affect the potential.  Consequently, the particle error is due only to collisions and time integration, and the potential error is due only to the basis functions.  In Figures~\ref{fig:uncoupled_coll_l2} and~\ref{fig:uncoupled_coll_linf}, the errors in the particles and the potential are $\mathcal{O}(h^2)$ for both norms.

Next, we consider two one-way couplings in Figures~\ref{fig:oneway_coll_l2} and~\ref{fig:oneway_coll_linf}.  For the first one-way coupling, we set $q/m=0$, such that the particles affect the potential but are not affected by the electric field.  The resulting error in the potential is due to the basis functions, finite particle sampling, and particle errors due to collisions and time integration; this error in the potential is shown in these figures.  For the other one-way coupling, we set $q=0$, such that the electric field affects the particles but the potential is not affected by the particles.  The resulting error in the particles is due to the collisions, time integration, and basis functions; this particle error is shown in these figures.  In Figures~\ref{fig:oneway_coll_l2} and~\ref{fig:oneway_coll_linf}, the errors in the particles and the potential are $\mathcal{O}(h^2)$ for both norms.  

Finally, we consider the full coupling of the field and the particles in Figures~\ref{fig:coupled_coll_l2} and~\ref{fig:coupled_coll_linf}.  The particle error is due to collisions, time integration, and all errors in the electric field.  The potential error is due to the basis functions, finite particle sampling, and all particle errors.  In Figures~\ref{fig:coupled_coll_l2} and~\ref{fig:coupled_coll_linf}, the errors in the particles and the potential are $\mathcal{O}(h^2)$ for both norms.

\subsubsection{Collisionless} 

For the collisionless case, we perform a similar study.  Once more, we consider two one-way couplings in Figures~\ref{fig:oneway_nocoll_l2} and~\ref{fig:oneway_nocoll_linf}.  The error in the potential is due to the basis functions, finite particle sampling, and particle errors due to time integration.  The error in the particles is due to the time integration and basis functions.  The errors in the particles and the potential are $\mathcal{O}(h^2)$ for both norms.

Figures~\ref{fig:coupled_nocoll_l2} and~\ref{fig:coupled_nocoll_linf} show the errors from the full coupling.  The particle error is due to time integration and all errors in the electric field.  The potential error is due to the basis functions, finite particle sampling, and all particle errors.  The errors in the particles and the potential are $\mathcal{O}(h^2)$ for both norms.

\section{Conclusions} 
\label{sec:conclusions}

In this paper, we presented our code-verification approaches for particle-in-cell simulations with and without collisions, and we derived expected convergence rates for the errors. 

For the particles, we incorporated the method of manufactured solutions into the equations of motion.  In doing so, we avoided modifying the weights and consequently the risks of incurring negative weights or modifying the collision algorithm.  By having known solutions for the particle positions and velocities, we were able to compute the error in these quantities directly instead of attempting to compute differences in distribution functions. 

To accommodate collisions, we averaged the outcomes from the collision algorithm, and we derived a corresponding source term.  We were able to compute the source term analytically by manufacturing the cross section and the anisotropy.  In addition to measuring the convergence of the particle positions and velocities, we measured convergence in the scattering-angle distributions as a supplementary method for detecting coding errors. 

We demonstrated the effectiveness of these approaches for multiple examples with and without coding errors, achieving the expected convergence rates with a single run per discretization. 

\section*{Acknowledgments} 
\label{sec:acknowledgments}

The authors thank Duncan McGregor, David Sirajuddin, and Thomas Smith for their insightful feedback. 
This article has been authored by employees of National Technology \& Engineering Solutions of Sandia, LLC under Contract No.~DE-NA0003525 with the U.S.~Department of Energy (DOE). The employees own all right, title, and interest in and to the article and are solely responsible for its contents. The United States Government retains and the publisher, by accepting the article for publication, acknowledges that the United States Government retains a non-exclusive, paid-up, irrevocable, world-wide license to publish or reproduce the published form of this article or allow others to do so, for United States Government purposes. The DOE will provide public access to these results of federally sponsored research in accordance with the DOE Public Access Plan \url{https://www.energy.gov/downloads/doe-public-access-plan}.

\clearpage
\appendix
\section{Collision Algorithm}

Algorithm~\ref{alg:coll} is based on the no-time-counter collision algorithm~\cite{bird_1994}.


\begin{algorithm}[H]
\caption{Binary Elastic Collision Algorithm (Same Species and Weight)}
\label{alg:coll}
\begin{algorithmic}[1]
\STATE \textbf{Input:} $N_p^\text{cell}$, $\mathbf{v}$, ${(\sigma g)}_\text{max}$, $\sigma(g)$, $w$, $\Delta t$, $\Delta V$

\vspace{1.0em}

\STATE \textbf{Output:} $\mathbf{v}$

\vspace{1.0em}

\STATE Compute the collision probability per pair: \qquad $\displaystyle P_{\text{coll}_\text{max}} = {(\sigma g)}_\text{max} w \Delta t /\Delta V$

\vspace{1em}

\STATE Compute the number of collision pairs: \qquad $\displaystyle N_{\text{pairs}} = \bigl\lfloor N_p^\text{cell} \bigl(N_p^\text{cell} - 1\bigr) P_{\text{coll}_\text{max}}/2 + \texttt{rand\_num()} \bigr\rfloor$

\vspace{1.0em}

\FOR{$k = 1$ \TO $N_{\text{pairs}}$}

    \vspace{1.0em}

    \STATE Select particle $p$: \qquad $\displaystyle p = \bigl\lfloor \texttt{rand\_num()} \cdot N_p^\text{cell} \bigr\rfloor + 1$
    
    \vspace{1.0em}

    \STATE Select particle $q$: \qquad $\displaystyle q = \bigl\lfloor \texttt{rand\_num()} \cdot N_p^\text{cell} \bigr\rfloor + 1$
    
    \vspace{1.0em}

    \WHILE{$p = q$}
        \vspace{1.0em}
        \STATE Update $q$: \qquad $\displaystyle q = \bigl\lfloor \texttt{rand\_num()} \cdot N_p^\text{cell} \bigr\rfloor + 1$
        \vspace{1.0em}
    \ENDWHILE
    
    \vspace{1.0em}

    \STATE Compute relative velocity magnitude: \qquad $\displaystyle g = |\mathbf{v}_p - \mathbf{v}_q|$
    
    \vspace{1.0em}
    
    \IF{$\texttt{rand\_num()}< \sigma(g) g/{(\sigma g)}_\text{max}$}
    
    \vspace{1.0em}

    \STATE Compute the azimuthal angle: \qquad $\displaystyle \epsilon = 2 \pi \cdot \texttt{rand\_num()}$
    
    \vspace{1em}
    
    \STATE Compute the polar angle (e.g.,~\eqref{eq:chi_iso}, \eqref{eq:vss}, \eqref{eq:fpiv_ms}): \qquad $\displaystyle \chi = F_{p_\chi}^{-1} (\texttt{rand\_num()})$
    
    %
    %
    %
    %
    %
    %
    %
    %
    
    \vspace{1em}

    \STATE Compute the post-collision relative velocity from $\mathbf{n}$~\eqref{eq:n}: \qquad $\displaystyle \mathbf{g}' = g \mathbf{n}$
    
    \vspace{1em}
     
    \STATE Compute the center-of-mass velocity: \qquad $\displaystyle \mathbf{v}_{\text{cm}} = (\mathbf{v}_p + \mathbf{v}_q)/2$
    
    \vspace{1em}

    \STATE Update the velocities of particles $p$ and $q$: \qquad $\displaystyle \mathbf{v}_p = \mathbf{v}_{\text{cm}} + \mathbf{g}'/2, \qquad \mathbf{v}_q = \mathbf{v}_{\text{cm}} - \mathbf{g}'/2$
    
    \vspace{1.0em}
    
    \ENDIF
    
    \vspace{1.0em}
\ENDFOR
\end{algorithmic}
\end{algorithm}

\bibliographystyle{elsarticle-num}
\bibliography{../../Gemma/quad_notes/quadrature_manuscript/quadrature.bib}


\end{document}